\documentclass[12pt,preprint]{aastex}

\slugcomment{}

\shorttitle{Clustering Properties of the GOODS LBGs}
\shortauthors{Lee et al.}
\begin{document}

\def\hh{\, h^{-1}}
\newcommand{\wth}{$w(\theta)$}
\newcommand{\mpc}{Mpc}
\newcommand{\xir}{$\xi(r)$}
\newcommand{\Lya}{Ly$\alpha$}
\newcommand{\Lyb}{Lyman~$\beta$}
\newcommand{\Hb}{H$\beta$}
\newcommand{\msun}{M$_{\odot}$}
\newcommand{\sfr}{M$_{\odot}$ yr$^{-1}$}
\newcommand{\dnsty}{$h^{-3}$Mpc$^3$}
\newcommand{\za}{$z_{\rm abs}$}
\newcommand{\ze}{$z_{\rm em}$}
\newcommand{\cmtwo}{cm$^{-2}$}
\newcommand{\nhi}{$N$(H$^0$)}
\newcommand{\degpoint}{\mbox{$^\circ\mskip-7.0mu.\,$}}
\newcommand{\halpha}{\mbox{H$\alpha$}}
\newcommand{\hbeta}{\mbox{H$\beta$}}
\newcommand{\hgamma}{\mbox{H$\gamma$}}
\newcommand{\kms}{\,km~s$^{-1}$}      
\newcommand{\minpoint}{\mbox{$'\mskip-4.7mu.\mskip0.8mu$}}
\newcommand{\mv}{\mbox{$m_{_V}$}}
\newcommand{\Mv}{\mbox{$M_{_V}$}}
\newcommand{\peryr}{\mbox{$\>\rm yr^{-1}$}}
\newcommand{\secpoint}{\mbox{$''\mskip-7.6mu.\,$}}
\newcommand{\sqdeg}{\mbox{${\rm deg}^2$}}
\newcommand{\squig}{\sim\!\!}
\newcommand{\subsun}{\mbox{$_{\twelvesy\odot}$}}
\newcommand{\et}{{\it et al.}~}
\newcommand{\er}[2]{$_{-#1}^{+#2}$}
\def\h50{\, h_{50}^{-1}}
\def\hbl{km~s$^{-1}$~Mpc$^{-1}$}
\def\ltsima{$\; \buildrel < \over \sim \;$}
\def\simlt{\lower.5ex\hbox{\ltsima}}
\def\gtsima{$\; \buildrel > \over \sim \;$}
\def\simgt{\lower.5ex\hbox{\gtsima}}
\def\arcs{$''~$}
\def\arcm{$'~$}
\newcommand{\wu}{$U$}
\newcommand{\wb}{$B_{435}$}
\newcommand{\wv}{$V_{606}$}
\newcommand{\wi}{$i_{775}$}
\newcommand{\wz}{$z_{850}$}
\newcommand{\hmpc}{$h^{-1}$Mpc}
\title{The Large--scale and Small--scale Clustering of \\ Lyman--Break Galaxies
 at $ 3.5 \leq z \leq 5.5$\\ from the GOODS survey}
\author{Kyoung-Soo Lee\altaffilmark{1,2}, 
Mauro Giavalisco\altaffilmark{2},
Oleg Y. Gnedin\altaffilmark{3}, 
Rachel S. Somerville\altaffilmark{2,4},
Henry C. Ferguson\altaffilmark{2}, 
Mark Dickinson\altaffilmark{5},
Masami Ouchi\altaffilmark{2,6}}
\altaffiltext{1}{Johns Hopkins University, Baltimore, MD 21218}
\altaffiltext{2}{Space Telescope Science Institute, Baltimore, MD 21218}
\altaffiltext{3}{Ohio State University, Columbus, OH 43210}
\altaffiltext{4}{Max-Planck-Institut f\"ur Astronomie, Heidelberg, Germany}
\altaffiltext{5}{National Optical Astronomy Observatory, Tucson, AZ 85719}
\altaffiltext{6}{Hubble fellow}

\begin{abstract}
We report on the angular correlation function of Lyman--break galaxies (LBGs)
at $z\sim4$ and $\sim5$ from deep samples obtained from the Great
Observatories Deep Origins Survey (GOODS).  As for LBGs at $z\sim3$, the
shape of \wth\ of the GOODS LBGs is well approximated by a power--law with
the slope $\beta\approx0.6$ at angular separation $\theta \geq 10$ arcsec. The
clustering strength of $z\sim4$, $5$ LBGs depends on the rest--frame UV
luminosity similar to that of $z\sim3$ ones, 
with brighter galaxies more strongly clustered than fainter ones.
At smaller separations, \wth\ significantly exceeds the
extrapolation of the large--scale power--law fit, implying enhanced spatial
clustering on comoving scales $r\leq 1$ Mpc.  We also find that bright LBGs
statistically have more faint companions on scales $\theta\la20$ arcsec than
fainter ones. The enhanced small--scale clustering is very likely
due to sub--structure, reflecting multiple galaxies within the same massive halos.
A simple model for the halo occupation distribution along with the halo mass
function in a $\Lambda$CDM cosmology, reproduces well the observed \wth.
The scaling relationship of the clustering strength with volume density and
with redshift is quantitatively consistent with that of CDM halos. If we 
associate LBGs with dark matter halos that have the same clustering strength,
this luminosity dependence of \wth\ implies a close correlation between the
halo mass and the star formation rate. A comparison of the clustering strength
of three samples of equal luminosity limit at {\it z} $\sim$ 3, 4 and 5 shows
that the LBGs at $z\sim5$ are hosted in halos about $5$ -- $10$ times less massive
than those at $z\sim 3$ -- 4, suggesting that star--formation was more
efficient at $z\sim 5$.
\end{abstract}


\section{Introduction}
In the current theoretical framework of galaxy evolution, the observed spatial
clustering of galaxies is interpreted as reflecting the clustering of the dark matter
halos that host them. Galaxies of different luminosity, spectral type,
morphology, etc. are observed to have, in general, different clustering
properties, because the specific criteria adopted to select them also select
different ``types'' of halos. For example, because more massive halos have 
stronger spatial clustering than less massive ones, galaxies selected by
criteria that also select the high--end of the halo mass function will be
observed to be more clustered than galaxies whose selection criteria
correspond to less massive halos. Similarly, the apparent ``evolution'' of the
spatial clustering of galaxies as a function of redshift is interpreted as
the result of the specific ``mapping'' of the selection criteria of the
galaxies into a mass range of the halo mass function and of the evolution of
the clustering of the halos, which is driven by gravity. 

This dependence of galaxy clustering on the selection criteria and how they
map into properties of the halos is at the same time bad and good news. It is
bad news because it makes it very difficult to use galaxies to infer anything
about the evolution of dark matter, and hence about cosmology, unless one knows
exactly how to go from galaxies' properties to halos' properties, including
taking into account the ``astrophysical'' evolution of the former, a very
complex problem.  It is good news because, at least for certain type of
galaxies and in some simple cases, one can expect to learn something abut the
physical relationship between certain properties of the galaxies and those of
the underlying halos, possibly testing some of the key ideas behind our
understanding of galaxy evolution.

One important case where the relationship between galaxies and halos is being
successfully explored by means of the analysis of the spatial clustering is
that of the Lyman--break galaxies. These are star--forming galaxies whose
rest--frame UV spectral energy distribution is not severely obscured by dust.
This type of source is very effectively selected by means of their
broad--band UV colors, constructed from around the 912 \AA\ Lyman limit, the
Ly$\alpha$ forest region (pronounced in broad--band photometry at $z\ga 2$),
and the otherwise relatively featureless continuum redward of the Ly$\alpha$
line (e.g. Steidel \& Hamilton 1993; Steidel et al. 1995; Madau et al. 1995).
Using optical passbands, where one can take advantage of the low sky and of
the sensitivity and large area coverage of CCD imagers and spectrographs, the
technique is effective at redshifts $2.5\la z\la 6$. Large surveys of
Lyman--break galaxies have been conducted over the past ten years in this
redshift range, resulting in large and well controlled samples, including large
spectroscopic ones (e.g. Steidel et al. 1999; 2003; Giavalisco et al. 2004;
Dickinson et al. 2004; Sawicki \& Thompson 2005; Vanzella et al. 2005;
E. Vanzella et al., in preparation). These surveys are also well suited for
clustering measures. 

Lyman--break galaxies have been found to have relatively strong spatial
clustering, with a correlation length comparable to that of present--day bright
spiral galaxies (Giavalisco et al. 1998; Giavalisco \& Dickinson 2001;
Porciani \& Giavalisco 2002; Adelberger et al. 2005). This has been
interpreted as evidence that they are biased tracers of the mass density field
(Giavalisco et al. 1998; Adelberger et al. 1998; Giavalisco \& Dickinson
2001; Bagla 1998). This means that these galaxies are preferentially hosted in
the most massive dark--matter halos
(Giavalisco \& Dickinson 2001; Foucaud et al. 2003; Adelberger et
al. 2005), which have stronger spatial clustering than less massive halos
(see, e.g. Mo \& White 1996). A key piece of evidence linking the mass of the
halo to physical properties of the galaxies came with the discovery that the
strength of the spatial clustering of Lyman--break galaxies is a function of
their UV luminosity, with brighter galaxies having larger correlation length
(Giavalisco \& Dickinson 2001; Foucaud et al. 2003; Ouchi et al. 2004;
Adelberger et al. 2005; Allen et al. 2005). 
In the halo interpretation, this is explained if
brighter galaxies are, on average, hosted in more massive halos. Since the
halo mass function is relatively steep, this also implies that the
statistical relationship between mass and UV luminosity must be characterized
by a correspondingly small scatter (Giavalisco \& Dickinson 2001). Because the
UV luminosity of Lyman--break galaxies is powered by star formation, this in
turn means that the mass and the star--formation rate are relatively tightly
correlated or, in other words, that the halo mass is a primary parameter that
controls the star--formation activity of the Lyman--break galaxies as expected
in galaxy formation models.

The next logical step is to try to derive the relationship between mass and
star--formation rate, thus testing the basic ideas behind the physical models
of star--formation in young galaxies (e.g. White \& Rees 1978). Furthermore,
the dependence of the clustering strength on the UV luminosity (clustering
segregation hereafter) contains direct information on the scaling relationship
between the halo clustering strength (bias) and the volume density once 
the galaxy luminosity function is known. Since the shape of
this relationship depends on the shape of the power spectrum, the observed
clustering segregation can be used to test key predictions of the CDM theory
(Giavalisco \& Dickinson 2001). To do this, however, information on the halo
sub--structure, or the halo occupation distribution function (HOD), is also
required (e.g. Bullock et al. 2001, Berlind \& Weinberg 2002). This function
describes the likelihood that one halo contains more than one galaxy and how
this depends on its mass. Once the HOD and the clustering segregation have
been measured, then the relationship between mass and UV luminosity
(star--formation rate) can be constrained, and a successful model will have to
simultaneously reproduce the observed clustering segregation and UV luminosity
function.

In principle, the presence of sub--structure and the HOD function can be
constrained by the shape of the angular correlation function and, as we will
see later, by the statistics of close pairs of galaxies. In practice, this
requires measures with relatively high S/N from large and faint samples,
unavailable until recently, since most of the secondary halo--companion
galaxies have lower mass than the central galaxy and are thus less luminous on
average.  Furthermore, large samples covering a sufficiently wide dynamic
range in luminosity are necessary to provide a robust measure of the shape of
the clustering segregation.  Giavalisco \& Dickinson (2001) first reported the
detection of the clustering segregation and measured its shape but with low
S/N. Using a deep sample of LBGs at $z\sim 4$ and 5 obtained with the Subaru
telescope and the Suprime camera, Hamana et al. (2004) reported evidence that
the angular correlation function of LBGs deviates from a power--law 
at small angular scales and derived the HOD function to explain the
deviation. Using the largest sample of LBGs to date at $z\sim 3$, Adelberger
et al. (2005) measured the shape of the angular correlation function at large
angular separations with high S/N. Their sample, however, is not deep enough
to include halo companion galaxies in appreciable number and detect the effect
of sub--structures; they could only set an upper limit of 5\% to the number of
galaxies in common halos.

In this paper we present a study of the clustering properties of Lyman--break
galaxies at $z\sim 4$ and $\sim 5$ at faint flux levels. We use two deep and
relatively large samples obtained with {\it HST} and ACS during the program of
multi--band imaging of the Great Observatories Origins Deep Survey (GOODS). In
fact, these ACS samples of \wb\ and \wv--band dropouts are
currently the largest and the most complete at their depth of such
galaxies\footnote{Analogous samples have also been obtained during the Hubble
Ultra Deep Field survey, which covers a small portion of the GOODS southern
field (CDF--S) with identical passbands as the primary GOODS survey. While
significantly deeper than the GOODS samples, the HUDF ones are much smaller in
size because of the minimal areal coverage of the HUDF.}. Since their size is
suitable to measure the angular clustering with high S/N, we have embarked on
a project to characterize the physical association between the activity of
star formation of the galaxies and the properties of the dark matter halos,
primarily the mass, thus providing quantitative tests to the CDM hierarchical
theory of galaxy formation. Here we present significantly improved measures of
the clustering segregation down to faint flux limits (\wz $\sim 27$), show direct
evidence of sub-structure in halos, and derive the corresponding HOD. We
postpone to a following paper the derivation of the relationship between mass
and star--formation rate.  All magnitudes in this paper are in the AB scale of
Gunn \& Oke (1975) and, when necessary, we use a cosmology with
$\Omega_m=0.3$, $\Omega_{\Lambda}=0.7$, $\sigma_8 = 0.9$, $\Gamma = 0.21$,
$H_0=100 h$ km s$^{-1}$ Mpc$^{-1}$ with $h=0.7$ and baryonic density
$\Omega_b=0.04$.

\section{The Data and The Samples}

Deep optical multi--band imaging with {\it HST} and ACS has been acquired as
part of the program of observations of the GOODS. These ACS data consist of a
mosaic of images in each of the \wb, \wv, \wi\ and \wz\ passbands for a total
of 3, 2.5, 2.5 and 5 orbits of integration time, respectively, in each of the
two GOODS fields. These two fields are centered around the Hubble Deep Field
North (HDF--N) and Chandra Deep Field South (CDF--S) and cover approximately
10$\arcmin$$\times$17$\arcmin$ each, for a total area coverage of about 0.1
square degree. To search for supernovae at high redshift (Riess et al. 2004),
the observations in the \wv, \wi\ and \wz\ bands were not done continuously,
but sequenced into ``epochs'' separated by about 45 days from each
other. Because the nominal roll of {\it HST} changes at a rate of about 1
degree per day, this observing strategy resulted in images acquired during
different epochs having position angle in the sky differing by integer multiple
of 45 degree. This gives the characteristic jagged edges of the GOODS mosaics
in the \wv, \wi\ and \wz\ bands. Giavalisco et al. (2004a) present an overall
description of the GOODS project, and we refer the reader to that paper for
additional details on the ACS observations.

The \wv, \wi\ and \wz--band mosaics presented in Giavalisco et al. (2004a,b)
only include the first 3 epochs of observations stacked together, while the
samples of Lyman--break galaxies at $z\sim 4$ and $\sim 5$ that we are about
to discuss here have been extracted from the version v1.0 reduction of the ACS
data, which include the full stack of the five epochs.  The \wb--band mosaic
discussed by Giavalisco et al. (2004a), however, already includes all the
data, since all the $B$--band observations were taken continuously. A complete
description of the v1.0 ACS images will be presented in an upcoming paper
(M. Giavalisco et al., in preparation).  Detailed information can also be found
in the GOODS website, at the link
{\tt http://www.stsci.edu/science/goods/v1\_release\_readme/h\_goods\_v1.0\_rdm.html}.

We now provide a brief summary of the data reduction and stacking process for
the version v1.0 data release. Differently from the version v0.5 public
release and from the mosaic discussed in Giavalisco et al. (2004a), for the
version v1.0 reduction we have re--processed all the data (including the
\wb--band images) using the best calibration and reference files available. We
have improved treatments of the geometrical distortions of ACS using updated
coefficients together with corrections for the velocity aberration distortion,
and we have also improved rejection of cosmic rays and other blemishes. We
have stacked the individual exposures together in each band using the
drizzling algorithm in two independent phases. In the first phase, images
taken in the same filters are sky-subtracted and drizzled onto a common pixel
grid at the instrument native scale (0.05 arcsec/pixel). Cosmic rays and
deviant pixels are identified during this process and flagged in mask files
specifically created for this purpose. Information on detector blemishes (hot
pixels, bad columns, etc.) in post--pipeline masks (which have also been
drizzled onto the same grid) is included in the new masks at this time. During
the second phase, the images and the mask files are blotted back to the
original positions, drizzled again onto a common astrometric grid with scale
0.03 arcsec/pixel, and stacked together. During this process corrections for
the ACS geometrical distortion are applied, cosmic rays flagged during the
previous processing block are masked out from the stack, and additional,
low-level cosmic rays and defects are identified and masked, too. 
The final stacks reach similar 1--$\sigma$ surface brightness limits for the 
north and south, which are listed in Table 1. 

Multi--band source catalogs, which are public and available at the Web site
listed above, have been created using the SExtractor package (Bertin \&
Arnouts 1996). We made the detections using the \wz--band mosaic, and then
used a variety of photometric apertures, including isophotal and a suite of
circular apertures of varying radius, as ``fixed apertures'' to carry out
matched photometry in all bands. We have measured the completeness of the
catalogs as a function of the size and magnitude of the sources through
extensive Monte Carlo simulations. In these simulations we have inserted a
very large number of model galaxies (169,100) into the ACS mosaic, after
convolution with the ACS PSF and after adding the appropriate Poisson and
detector noise. We used galaxies with both exponential and $r^{1/4}$ light
profiles with random ellipticity and orientation in the sky, with varying
apparent magnitude and size, and then retrieved them with identical procedures
as for the real galaxies. From these simulations we estimate that the catalogs
are $\approx 50$\% complete at \wz $\sim 26.5$ for sources with half--light
radius $r_{1/2}\le 0.2$ arcsec, and about 10\% complete at \wz $\sim 27$. We
have also run multi--band simulations for sources with a given input spectral
energy distribution (SED) to test completeness relative to color selection, as
we will discuss in a moment.

We have also used deep ground--based multi--wavelength imaging data to select
samples of Lyman--break galaxies at redshifts $z\sim 3$ (\wu--band dropouts)
in the CDF--S field, albeit in a larger area than the ACS fields. These data
were acquired as part of the program of ancillary observations of the GOODS
project. They are also described in Giavalisco et al. (2004a), and we refer
the reader to that publication for details. Here we will summarize the key
properties of these data.  The $B$ and $R$ bands were taken with the Wide
Field Imager (WFI) at the 2.2m MPG/ESO telescope, while the \wu--band
observations were carried out with the MOSAIC II Camera at the CTIO 4m
telescope. The stacked image in each band covers a contiguous region of
approximately 0.4 degree$^2$.  Some parameters of these ground--based data sets are detailed
in Table 2. \\

Samples of star--forming galaxies at $z\sim$ 3, 4 and 5 were photometrically
selected using the Lyman Break technique (e.g. Steidel et al. 1999; Madau et
al. 1996; Giavalisco et al. 2004b.; see also Giavalisco 2002 for a
review). The technique and its application to clustering studies (Giavalisco
et al. 1998; Adelberger et al. 1998; Giavalisco \& Dickinson 2001) have been
extensively discussed in the past, and we refer the reader to the cited
literature.

Lyman--break galaxies (LBGs) at $z\sim 4$ and $\sim 5$,
\wb\ and \wv--band dropouts, respectively, were extracted from the GOODS v1.0
r1.1 catalogs using the same color selection criteria described by Giavalisco
et al. (2004b), which we report here for convenience. We defined the 
\wb\ and \wv--band dropouts as 
\begin{eqnarray}
(B_{435} - V_{606}) &\geq& 1.2 + 1.4 \times (V_{606}-z_{850}) \wedge  \nonumber \\
(B_{435} - V_{606}) &\geq& 1.2 \wedge V_{606}-z_{850} \leq 1.2,  \nonumber
\end{eqnarray}
and
\begin{eqnarray}
(V_{606}-i_{775}) &>& 1.5+0.9 \times (i_{775}-z_{850}) \vee  \nonumber \\
(V_{606}-i_{775}) &>& 2.0 \wedge (V_{606}-i_{775}) \geq 1.2 \wedge  \nonumber \\ 
(i_{775}-z_{850}) &\leq& 1.3, \nonumber 
\end{eqnarray}
where the symbols $\vee$ and $\wedge$ are the logical ``OR'' and ``AND''
operators, respectively. In both cases we have limited the apparent magnitude
of the samples to \wz $\leq 27.0$ for completeness. We have also visually 
inspected each
galaxy included in the samples and removed objects with stellar morphology 
and obvious spurious detections, such as diffraction spikes. Down to
the adopted flux limit of \wz $ \leq 27.0$, the ACS samples include 2463 
\wb--band dropouts and 878 \wv--band dropouts.

Lyman--break galaxies at $z\sim 3$ were selected using
\begin{eqnarray}
(U - B) &\geq& (B - R) + 0.6 \wedge \nonumber \\
(U - B) &\geq& 0.9 \wedge (B - R) \leq 2.5 \nonumber 
\end{eqnarray}
This ground--based sample includes 1609 galaxies in the central region of the
field which was used for the clustering analysis down to a flux limit of $R$
$\leq $ 25.5.  
The corresponding surface density values of our samples down to the above
mentioned flux limits are 
$\Sigma$ = 2.7, 7.7,
2.7 galaxies/arcmin$^2$ for the \wu, \wb\ and \wv--dropouts, respectively.

Contamination from galactic stars is nearly negligible for the \wb\ and
\wv--band dropouts because they were morphologically culled taking advantage
of the high angular resolution of the GOODS ACS data. For the ground-based
sample such culling is not possible; however stellar contamination in high
galactic latitude fields such as GOODS has been found to be $\approx 4$\% in
previous spectroscopic surveys (Steidel et al. 1998). We also impose a minimum
signal to noise ratio (S/N $\geq$ 10) in order to filter out spurious sources.
This also helps (relatively) homogeneous detection of star-forming
galaxies up to a certain magnitude limit when dealing with data 
produced with several separate pointings with overlaps. Even at
the cost of losing some fraction of the faintest LBGs, it is desirable for
clustering studies to avoid an artificial clustering signature induced by
interlopers and inhomogeneity of the data. 

\section{Simulations}
At $z\sim 3$ the selection of Lyman--break galaxies (\wu--band dropouts) has
been characterized in great detail thanks to the systematic spectroscopic
identification of thousands of candidates (e.g. Steidel et al. 2003). In
particular, this extensive body of work has shown that the Lyman--break
technique is very efficient, with a relatively low rate of contamination from
low--redshift interlopers, and has yielded the redshift distribution function
$N(z)$ associated with a given set of color selection criteria.  While
spectroscopic identifications have been made of Lyman--break galaxies at
$z\sim 4$ and $\sim 5$ (Steidel et al. 1999; Giavalisco et al. 2004),
including a number presented in this work (see later), the spectroscopic
samples still remain too small for accurate measures of the redshift
distribution function associated with a specific color selection. Fortunately,
as the $z\sim 3$ case has demonstrated (Steidel et al. 1999), Monte Carlo
simulations are very effective in providing robust estimates of $N(z)$, which
are needed to derive the real--space clustering from the angular one and to
measure the galaxies' volume density.

The simulations consists of creating artificial LBGs over a large redshift
range (we used $2\le z\le 8$) with assumed distribution functions of UV
luminosity, SED, morphology and size, inserting them in the real images with
random orientations and inclinations and with the appropriate amount of photon
noise, detecting them, and measuring their ``observed'' photometry and
morphology as if they were real galaxies. We have adjusted the input
distribution functions of UV luminosity (UV luminosity function), size and SED
so that the observed distribution functions of apparent magnitude, UV colors
and size of the simulated galaxies best reproduce the same quantities of the
real galaxies.  Specifically, we used galaxies with exponential and $r^{1/4}$
light profile in equal proportions and size extracted from a log--normal
distribution function, as described by Ferguson et al. (2004).  This method
was first used by Giavalisco et al. (2004b) and Ferguson et al. (2004); its
application to the measure of the UV luminosity function of LBGs will be
further discussed in an upcoming paper (M. Giavalisco et al., in preparation).

The average redshifts obtained from the simulations are {\it z} = 3.2 
$\pm$ 0.3, 3.8 $\pm$ 0.3, 4.9 $\pm$ 0.3
 for the \wu, \wb\ and \wv--band dropouts, respectively. 
Figure 1 shows the redshift distribution $N(z)$ of simulated LBGs together with 
that of a subset of LBGs with spectroscopic redshifts.

\section{Measuring the clustering properties of Lyman--break Galaxies}

We now discuss the measure of angular and real--space clustering of LBGs and other
clustering properties that we have carried out from our samples. The primary
measure is that of the angular correlation function \wth, which we
deproject with the Limber transform using the redshift distribution function
estimated from the simulations (but see later for a discussion on our
spectroscopic observations) to derive the real--space correlation length, in the
power--law approximation. In our faint samples, the angular correlation
function is very well approximated by a power law at large angular
separations, roughly $\theta>10$ arcsec, but it exceeds the extrapolation
of the large--scale power--law fit at small scales. To understand the nature
of this excess clustering, which is not observed in bright samples, either our
own \wu--band dropout sample or even larger ones (e.g. Adelberger et al. 2005),
we have carried out an analysis of the mean number of close neighbors around
both bright and faint galaxies. We have used the halo occupation 
distribution (HOD) formalism to predict the shape of \wth\ and compare it with
our measure.

\subsection{The angular correlation function}

The inversion of the angular correlation function is a robust method to derive
the real--space correlation length if the redshift distribution function is
well known (see Giavalisco et al. 1998; Giavalisco \& Dickinson 2001;
Adelberger et al. 2005). The measure of \wth\ is relatively straightforward;
however some care is required in the analysis of the random errors, and, above
all, of systematic errors, since they can significantly bias the result. Given
the relatively small area covered by the GOODS fields and the way the
observations have been obtained, the two most significant sources of
systematic errors are cosmic variance (the integral constraint bias) and
sensitivity fluctuation in the survey, which can mimic a spurious clustering
signal. These will be discussed separately in dedicated subsections.

We have estimated the angular correlation function using the estimator
proposed by Landy and Szalay (1993): 
\begin{equation}
w_{obs}(\theta) = \frac{DD(\theta)-2DR(\theta)+RR(\theta)}{RR(\theta)}, 
\end{equation}
where $DD(\theta)$ is the number of pairs of observed galaxies with angular
separations in the range $(\theta-\delta \theta /2, \theta+\delta \theta /2)$,
$RR(\theta)$ is the analogous quantity for random catalogs with the same
geometry as the observed catalog and $DR(\theta)$ is the number of
random galaxy cross pairs. 
We have chosen logarithmic binning of the angular separations
to provide adequate sampling at small scales while avoiding excessively
fine sampling at large scales which would have resulted had linear binning
been adopted. We have repeated the measures using 4 different (logarithmic) binsizes for each sample.

To account for the random errors, we have estimated the error bars for each
angular bin by bootstrap resampling of the data (e.g. Benoist et al. 1996,
Ling, Barrow \& Frenk 1986). We have computed values of \wth\ from 100
randomly selected subsets of galaxies and then used the standard deviation of
\wth\ as our primary error estimate. This method, however, only accounts
for random errors appropriate to our sample size, and does not include the
 uncertainty due to the finite size of our samples, because we
use the observed galaxies for the error estimate. In other words, the
bootstrap method fails to account for fluctuations on scales larger than our
survey. We have used Monte Carlo simulations to estimate these fluctuations
of \wth\ due to cosmic variance and validate our accounting of random
errors. We have created a large number of realizations (50) of GOODS samples
at $z\sim 4$ and $\sim 5$ extracted from a population with the same
large--scale clustering properties (i.e. surface density and angular
correlation function--- see Porciani \& Giavalisco 2002 for a description of
the method) as the observed galaxies. Since the goal of this test is to estimate the
uncertainty induced in the observed \wth\ by the large--scale fluctuations of
the density field of the parent population of galaxies from which are samples
are extracted from, neglecting the small--scale clustering is
inconsequential. Each realization simulates an area of the sky of 1
degree$^2$ within which we have selected the mock ``GOODS'' samples of
LBGs. We have measured \wth\ in each sample, and derived the standard 
deviation of the 50 measures at each angular bin and used it as an estimate of
the error bar, finding consistent results with the bootstrap method. 

To test the dependence of the angular clustering on the luminosity of the
galaxies, we have measured the correlation function for various magnitude
cuts in each sample. For the \wb\ and \wv--band dropout samples we have
defined sub--samples at \wz $\le 27.0$, 26.5 and 26.0; for the \wu--band 
dropouts we have used $R\le 25.5, 24.5$ and $24.0$. In what follows, we will
denote each sub--sample as B(V)270, 265, 260 and U255, 245, 240, respectively.
The number of galaxies included in each sub--sample is 2463, 1410 and 858 for the
\wb--band dropouts, 878, 407 and 218 for the \wv--band dropouts, and
1609, 1050 and 554 for the \wu--band dropouts. 

\subsection{Correction for sensitivity variations}

A possible source of systematic error in surveys such as GOODS, which consist
of either mosaics of images (the ACS images) or images taken with CCD mosaics
(the ground--based images), is fluctuations of sensitivity across the surveyed
area. Such fluctuations, which typically occur on scales that are a fraction
of the linear size of the survey, can be sufficiently strong to mimic the
presence of clustering and bias the measures of \wth.

To estimate this effect in LBGs samples is not straightforward, because these
are color--selected as well as flux-limited, and thus the final selection
function is a nontrivial function of sensitivity variations in all the
different passbands used in the color selection as well as of their respective
limiting magnitudes. The best way to take all of these factors into account is
to use Monte Carlo simulations. We have generated over 100,000 artificial \wu\
and \wb--band dropouts and placed them at random positions in each of the \wu,
$B$, $R$ and \wb, \wv, \wi, \wz\ images of the ground--based and ACS survey,
respectively (see above for a description of the simulation technique). We
have subsequently ``detected'' the artificial galaxies and subjected them to
the same color selection as if they were real \wu\ or \wb--band dropouts. To
the extent that this process is a faithful representation of the overall
selection function of the survey, the spatial distribution of these
color--selected mock galaxies will reflect various systematics including the
sensitivity variations across the field, the effects of holes due to saturated
sources or large seeing (of our ground--based mosaics), to the applied
color--selection and so forth.

The correction factor
can be estimated by measuring the angular correlation function,
$w_{sim}(\theta)$, against a purely random distribution with the same geometry
(this should be essentially zero for perfectly homogeneous data). Since
\wth\ is the excess probability of finding a pair, our previous
measurement can be corrected as
\begin{equation}
1+w_{corr}(\theta) = \frac{1+w_{raw}(\theta)}{1+w_{sim}(\theta)}
\end{equation}
where $w_{raw}$ is the measurement using the LS estimator, $w_{sim}$ is the
correlation function of artificial galaxies and $w_{corr}$ is the corrected
angular correlation function of the dropouts.  While the correction is
non-negligible at small scales for the \wu--dropouts, it has negligible effect
on the ACS data, hence the correction is applied only to the \wu--dropout
measurement. The subscript of $w_{corr}(\theta)$ will be dropped hereafter
unless stated otherwise.

\subsection{Integral Constraint}
Due to the finite size of any given survey, the correlation function measured
from a sample is
underestimated by a constant known as the integral constraint, referred to as
$IC$ hereafter:
\begin{equation}
\omega_{true}(\theta) = \omega_{obs}(\theta) + IC
\end{equation}

Since the measure of \wth\ is underestimated by the same amount on all scales, this also
leads to an overestimate of the correlation slope, $\beta$. This can be
estimated by doubly integrating $\omega_{true}$ over the survey area (Roche 
\& Earles 1999).
\begin{eqnarray}
IC &=& \frac{1}{\Omega^2} \int_1 \int_2 \omega_{true}(\theta) d\Omega_1 \Omega_2 \nonumber\\
   &=& \frac{\Sigma_i RR(\theta_i) \omega_{true}(\theta_i)}{\Sigma_i RR(\theta_i)} = \frac{\Sigma_i RR(\theta_i) A_\omega \theta_i^{-\beta}}{\Sigma_i RR(\theta_i)}
\end{eqnarray} 

Note that the integral constraint depends on both the size of a given survey
($\Omega$) and the galaxy power spectrum (approximated by a power--law ; $A_w$
and $\beta$). However, once the correlation slope $\beta$ is fixed, $IC/A_w$
only depends on the size and the shape of the survey area.  Therefore, one
needs to make robust estimate of $\beta$ from the observational data, then for
a fixed $\beta$, $IC/A_w$ can be estimated.  We calculate the $IC$s from
random catalogs generated over a field with the same size and the geometry as
our survey area with a range of $A_\omega$ and $\beta$. For each set ($A_w$,
$\beta$) we calculate $IC$ value using Equation (4) and compute the $\chi^2$ of
the measured $w_{obs}(\theta)$.  Once $\beta$ and $IC/A_w$ are robustly
measured from the full sample, for other flux--limited samples, $A_w$ is the
only parameter that needs to be fitted as $w_{model} = A_w
(\theta^{-\beta}-(IC/A_w)_0)$ provided that $\beta$ does not change
significantly with samples of different flux limits.  This helps to make
reliable measures of the amplitude $A_w$ for sub--samples that have lower
signal--to--noise than the full sample.

For the B270 sample, we find that a set ($A_w$, $\beta$, $IC$) = (0.31, 0.6,
0.012) best describes the measures, however, it should be noted that there is
degeneracy between $A_w$ and $\beta$. In other words, there is a range of
acceptable $A_w$s and $\beta$s for a fixed $IC$ value even for high S/N data,
and hence one needs to somewhat arbitrarily fix the slope within the given range.
We fix $\beta = 0.6$ and $IC/A_w = 0.039$ for all \wb\ and \wv--band dropout samples. 
Similarly, for the
\wu--band dropouts, we find ($A_w$, $\beta$, $IC$) = (0.50, 0.6, 0.012) thus
$IC/A_w$ = 0.024.  Note that $IC/A_w$ is much smaller for the \wu--band
dropouts due to the larger field size. On the other hand, the $IC$ values for the
full sample of the \wb\ and \wu--band dropouts are comparable, because the
faint ACS \wb\ and \wv\ dropout samples are less clustered
(i.e. smaller $A_w$) than the brighter ground--based sample.  Using this
method, we find $IC$ = $0.012$, $0.024$, $0.031$ for the U255, U245 and U240 sample, 
$IC$ = $0.012$, $0.020$, $0.027$ for the B270, B265 and B260 sample and
$IC$ = $0.024$, $0.031$, $0.043$ for the V270, V265 and V260 sample, respectively.

When the slope is allowed to vary, the procedure becomes iterative because the
derived $IC$ depends on \wth\ and vice versa, and one cannot conveniently set
$IC/A_w$ to be a constant for the fitting.  As Adelberger et al. (2005)
pointed out, this is often very unstable regardless of the initial
guess. Therefore, as a consistency check we estimated the $IC$ using an alternate
method suggested by Adelberger et al. (2005). Their method takes advantage of
the fact that matter fluctuations, $\sigma_{CDM}^2$ within a given survey volume
are in the linear regime and can be estimated from the power spectrum with
an appropriate window function that accounts for the geometry of the survey
volume (see their Equation 20). By definition, $IC$, the variance of density
fluctuation of galaxies is $IC \approx$ $b^2$ $\sigma_{CDM}^2$.  We estimate
the galaxy bias, {\it b}, from the correlation function itself by using the
relation {\it b} = $\sigma_{8,g}/\sigma_8(z)$, where $\sigma_{8,g}$ is the
galaxy variance in spheres of comoving radius 8 $h^{-1}$ Mpc (Peebles 1980,
Equation [59.3]):
\begin{equation}
\sigma^2_{8,g} = \frac{72(r_0/8h^{-1} \mpc)^\gamma}{(3-\gamma)(4-\gamma)(6-\gamma)2^\gamma}
\end{equation}
where $r_0$ is the correlation length inferred from the angular correlation
function (see next subsection), $\gamma = \beta + 1$ and $\sigma_8 (z)$ is the
linear matter fluctuation on the same scale extrapolated from $\sigma_8(0)$ =
0.9.

When using this method, we find values of both $IC$ and $\beta$ that are consistent with the
previous method, but the $IC$ values are slightly smaller in most cases.  For
example, for the B270 sample, we find $IC$ = 0.010 instead of 0.012, for B265
0.018 instead of 0.020, which makes little difference in the
fitted $A_w$ or $\beta$ values. This may imply that the generic correlation
function slope $\beta$ may be slightly steeper than the fiducial value 0.6 for
some sub--samples as can be seen in Table 3. For a fixed $r_0$, $\sigma_{8,g}$
declines with the slope $\gamma$ at $1.5<\gamma <2.0$. However, for a fixed
$\gamma$ (or $\beta$), both methods are in good agreement.

Since the $IC$ is \wth--dependent, power--law fitting and correcting for the
$IC$ are in fact simultaneous processes. In each iteration, only the
$w_{obs}(\theta)$ measurement with $\theta > 10$ arcsec is used for the power--law
fit (to probe large scale structure) because at $\theta < 10$ arcsec there may
be significant contribution from highly nonlinear small-scale
clustering. Figure 2 shows our results for the full \wb--band (left panel)
 and \wv--band (right panel)
dropout sample together with the best power--law fit. Clear departure 
from a pure power--law is apparent in both cases on small scales ($\theta < 10$ arcsec).

We fit the data to a linear function in logarithmic space. 
To estimate 1$\sigma$ confidence intervals for the parameters $A_w$
and $\beta$, we carried out a large ensemble of random realizations of the
measured \wth\ assuming normal errors, and calculated best--fit parameter
values for each of these synthetic data sets. Because we select randomly among
the \wth\ measures of different binsizes for each realization, the peak values 
of $A_w$ and $\beta$ may be slightly different from those derived from the
individual \wth\ measures mentioned in the previous subsection.
The fitted values of $A_w$ and $\beta$ are provided in Table 3. 

\subsection{The Real--Space Correlation Function $ \xi (r)$ }

We derive the real--space correlation function $\xi$(r) by inverting \wth\
using the Limber transform (Peebles 1980).  If the real--space correlation function has the
form $\xi(r)=(r/r_0)^{-\gamma}$, the angular correlation function also has to
be a power--law, $w(\theta) = A_w \theta ^{-\beta}$ where $\beta = \gamma - 1$
and $A_w$ is related to $\xi(r)$ as,
\begin{eqnarray}
A_w = Cr_0^{\gamma}\int F(z) D^{1-\gamma}_{\theta} (z)N(z)^2g(z) dz
\times [\int N(z)dz]^{-2} \nonumber
\end{eqnarray}
where $D_{\theta}$ is the angular diameter distance and $N(z)$ is the redshift
selection function derived from the simulations.
\begin{eqnarray}
g(z) &=& \frac{H_0}{c}[(1+z)^2 (1+\Omega_0z+
\Omega_{\Lambda}((1+z)^{-2}-1))^{1/2}] \nonumber \\
C &=& \sqrt{\pi} \frac{\Gamma[(\gamma - 1)/2]}{\Gamma(\gamma /2)}.\nonumber
\end{eqnarray}

We have used the same $N(z)$ for all the different limiting magnitudes since
the simulations indicate little change in the redshift distribution for each case. 
Note that $A_w$ and $\beta$ are obtained from the measures at $\theta > 10$ arcsec to be
representative of large--scale clustering. It is interesting, however, that
if we include the data points at $\theta < 10$ arcsec for the fit and 
allow $\beta$ to vary, 
both $A_w$ and $\beta$ get overestimated by an amount that changes $r_0$ 
very little. For example, for the B270 sample, we obtained the $r_0$--value of 
$2.82\pm0.20$ $h^{-1}$ Mpc ($2.81^{+0.22}_{-0.21}$ $h^{-1}$ Mpc) 
when using the data points at $\theta > 3$ ($\theta > 10$) arcsec. 
Hence our estimates of $r_0$ are robust regardless of 
the fiducial value chosen to be considered as large--scale.

The correlation lengths $r_0$ derived using this method are listed in
Table~\ref{tbl-3}. Two different $r_0$ values derived from the fits with and
without the correlation slope $\beta$ fixed to a fiducial value 0.60, are
consistent with each other within the errors. In the following discussion, we
regard the former as our best $r_0$ values.
The best--fit $r_0$ values range from 2.8 $h^{-1}$ Mpc to 7 $h^{-1}$ Mpc
(comoving) depending on the median luminosity of the sub--samples. In each
redshift range, brighter samples have larger correlation lengths. We have
compared our measures with those from other groups for samples with similar
median luminosity and in the same redshift range generally finding good
agreement. At $z\sim 3$ Adelberger et al. (2005) found $r_0$ = 4.0 $\pm$ 0.6
$h^{-1}$ Mpc for galaxies with magnitude 23.5$\le\cal{ R}\le$25.5 ($\bar{z}$ =
2.9), in good agreement with our $R$ $\le$ 25.5 sample for which we found $r_0
= 4.0\pm0.2$ $h^{-1}$ Mpc.  At {\it z} $\sim$ 4 and 5, Ouchi et
al. (2004) reported $r_0 = 4.1\pm0.2$ and $5.9^{+1.3}_{-1.7}$ $h^{-1}$
Mpc for their LBG$z$4s ($i'\lesssim 26$) and LBG$z$5s ($z'<25.8$)
samples, respectively. Using an LBG template spectrum that reproduces the
median UV colors of the samples, we computed that the flux limit of Ouchi et
al. (2004) correspond to \wz$ \la 26.0$ and 25.8, for which we find $r_0$ =
5.1$^{+0.4}_{-0.5}$ and 5.30$^{+1.1}_{-1.0}$ $h^{-1}$ Mpc. While the
correlation lengths of the two samples of \wv--band dropouts are in good
agreement, we noticed that Ouchi et al. reported a lower value than ours for
the \wb--band dropout sample, although the two measures overlap at the
$\approx 1.3\sigma$ level.

\section{Results}

\subsection{Dependence of Clustering strength on Luminosity}

A number of authors (Giavalisco \& Dickinson 2002; Foucaud et al. 2003;
Adelberger et al. 2005; Ouchi et al. 2004) have reported the existence of the clustering
segregation with the UV luminosity for the LBGs, namely of the fact that the
galaxies with higher UV luminosity have stronger spatial clustering
(e.g. larger spatial correlation length) than fainter ones. We have examined
the dependence of the clustering strength of our samples on luminosity using
several apparent magnitude cuts constructed from the full samples. Apparent
magnitude cuts roughly correspond to absolute magnitude cuts because the
luminosity--distance changes only by $\approx$ 20\% over the redshift range
covered by each of our samples and the redshift selection functions are
relatively narrow.  Whether or not we fix the correlation slope to a fiducial
value of $\beta$ = 0.6 (Porciani \& Giavalisco 2002), we find that the
correlation amplitude $A_w$, and $r_0$ accordingly, increases with median
luminosity.

The correlation length of the \wb--band dropouts increases from
$2.8\pm0.2$ $h^{-1}$ to 3.7$\pm0.3$ $h^{-1}$ to
5.1$^{+0.4}_{-0.5}$ $h^{-1}$ Mpc with increasing luminosity of the samples
(see Tables 3). 

Note that the faintest sample of \wb--band dropouts, which is our faintest
sample at any redshift, reaching absolute magnitude $M_{1700}=-18.52$, also has
the smallest correlation length, $r_0=2.8$ $h^{-1}$ Mpc. A similar trend is
also found for the \wv--band dropouts. For a fixed absolute luminosity,
objects will be fainter at $z\sim$ 5 than their counterpart at $z \sim$ 4 by 0.6
-- 0.7 mag in \wz--band due to cosmological dimming. This suggests that the
correlation lengths for the \wb\ and \wv--band dropouts for a fixed absolute
UV luminosity are comparable (see Table 3 for details).

In agreement with previous results, we find that the \wu--band dropouts also follow
a similar trend. For the U255 sample ($R$ $\le$ 25.5) we measure a
correlation length $r_0=4.0^{+0.2}_{-0.2}$ $h^{-1}$ Mpc, in an excellent
agreement with the value reported by Adelberger et al. (2005), 4.0 $h^{-1}$
Mpc for roughly the same magnitude threshold ($\langle R - {\cal R}\rangle$
$\sim$ 0.03 at {\it z} $\sim$ 3 using our mean LBG spectrum). The correlation
length increases to $r_0=7.8^{+0.5}_{-0.6}$ $h^{-1}$ for $R \leq$ 24.0.
Interestingly, this is comparable to 8 $h^{-1}$ Mpc, the correlation length of
a sample of distant--red galaxies (DRGs) found at a slightly lower redshift range (Daddi et
al. 2003).

The left three panels of Figure 3 show \wth\ for the full sample (solid lines) and the
brightest sub--samples (dashed lines) for each dropout flavor together with best--fit
power--laws. The power--fits when $\beta$ is fixed to 0.6 are
shown for clarity. The right panels show the derived correlation lengths for
these samples.  All samples show a clear trend that brighter samples are
more strongly clustered in real space.

The sub--samples that we have used to measure the clustering segregation are
not independent from one another, because the galaxies that comprise the
brighter samples also belong to the faintest one thus diluting the weak
clustering signal from the faintest galaxies.
We have repeated the measures using two mutually independent samples that we
built by splitting the full \wb--band dropouts sample into two independent
magnitude bins, $26.3<$ \wz $\leq 27.0$ and \wz$ \leq 26.3$ with roughly equal
number of galaxies in each bin. For these two samples we have measured
$r_0=4.0\pm0.3$ and 2.3$^{+0.5}_{-0.4}$ $h^{-1}$ Mpc,
respectively. Note that the correlation length of the faint sample is now
smaller than $2.8\pm0.2$ $h^{-1}$ Mpc of the \wz $ \leq 27.0$ sample,
consistent with the luminosity segregation.

Recently, Norberg et al. (2001) and Zehavi et al. (2002) observed 
luminosity segregation in the optically selected galaxies in local universe
with high S/N.  
Note that a direct comparison between our results and the local
ones is not possible, since the UV luminosity is powered by star formation rate, 
while the optical luminosity is mostly powered by stellar mass. 
However, understanding how UV luminosity scales with the clustering strength, at 
both high and low luminosities,
will shed light on the relationship between halo mass and star formation rate.  
Unfortunately, our brightest magnitude cut made to the ACS sample
is \wz = 26.0, which is still more than 1 mag fainter than $m^*$ ($m^*$ is \wz
$\sim$ 24.7 and 24.8 at {\it z} = 4 and 5; M. Giavalisco et al., in preparation),
while the size of our U dropout samples are relatively small due to the shallow
depth of our ground--based data sets. 
We plan to extend our observation of
clustering segregation to the higher--luminosity regime at the same redshift
intervals ({\it z} $\sim$ 3 -- 5) in a follow--up paper (K.-S. Lee et al., in preparation). 

\subsection{Small--scale statistics}

Our deep ACS data also allow us to probe the existence of sub--structure in
the halos that host the most massive LBGs. A general feature of hierarchical
clustering is that more massive halos have higher probability than smaller
ones to host more than one galaxy (e.g. Berlind \& Weinberg 2002). If the
existence of the clustering segregation is due to the fact that LBGs with
higher UV luminosity are, on average, more massive than fainter ones, then
one prediction is that brighter LBGs should also show a tendency to have more
closely associated galaxies than fainter ones. Our deep ACS sample seems well
suited to carry out this test, because satellite galaxies are typically less
massive than the central galaxy of the halo (Berlind \& Weinberg 2002), and
because less massive galaxies apparently have fainter UV luminosity.  It is
likely that pairs, or multiplets, in which one or more members are
significantly fainter than the brightest one are under--represented in the
shallower ground--based samples. Unfortunately, the lack of systematic
redshift measures prevents us from identifying individual physical pairs. Our
measure will have to be statistical in character, and signal--to--noise will
be lost due to projection.

In our samples, pair statistics on small scales reflect not only
large--scale structure of dark matter (via projected close--pairs) but also
the physics of galaxy formation within given halos (via physical
close--pairs). In this section, we discuss small scale statistics of the ACS
selected samples only (\wb\ and \wv--band dropouts).

As shown in Figure 2, a significant departure from a power--law on small
scales is observed in our \wth\ measures, most pronounced at the scale of $\theta <
10$ arcsec. $10$ arcsec corresponds to roughly 0.3 -- 0.4 $h^{-1}$ Mpc comoving at
redshift 3 -- 5. Recently, Ouchi et al. (2005b) detected a similar small--scale
excess in their \wth\ measurement of $z\sim4$ galaxies selected from 
a deep wide--field survey in SXDF field (Ouchi et al. 2004). 
Steepening on small scales seems to be present in all
samples, which we verify by fitting the \wth\ measurement of each sample to 
a power--law when the data
points with separations $\theta \leq 10$ arcsec are included, in which case
we find that the slope is substantially steeper. The slopes of the power--law
fits in both cases are shown in Figure 4 for the \wb\ and \wv--band
dropouts.

This is consistent with the theoretical expectation that there are two
separate contributions to \wth\ or $\xi(r)$, namely a one--halo term and a
two--halo term, the origins of which are fundamentally different in
nature. The two--halo term reflects the spatial distribution of host dark
halos by counting galaxy pairs that belong to two distinct halos. This term
dominates large--scale behavior of the correlation function and vanishes on
small scales due to the halo exclusion effect, namely two halos cannot coexist
closer than their typical size.  In contrast, the one--halo term accounts for
galaxy pairs sharing the same halo and is dominant on small scales. This
reflects how galaxies populate halos.  In this context, the single power--law
nature of the correlation function seen in the local universe is somewhat
coincidental because it requires a smooth transition between these two terms
to result in pure power--law which is possible only for a certain distribution
of galaxies therein. Berlind \& Weinberg (2002) argued that the contribution
of galaxy pairs in massive galaxy clusters is critical for a smooth transition
between these two terms. Clusters must be rare at high redshift, hence this
steepening should be more prominent. This is also observed in high--resolution
N--body simulations. For example, Kravtsov et al. (2004) measured the
correlation function of halos above a given mass threshold identified from
such simulations at various cosmic epochs. They found that at $z=0$ the halo
correlation function can be fairly well approximated by a power law at all
probed scales. However, at higher redshifts, steepening occurs on small
scales, at progressively smaller scales with higher redshifts.  As a result,
power--law fits using the range of scales 0.1 -- 8 $h^{-1}$ Mpc give
systematically steeper values of $\beta$ compared with the fits over the range
$\sim 0.3$ -- $8h^{-1}$ Mpc. In fact, Figure 4 shows a similar behavior (in
the angular domain) as Figure 12 of Kravtsov et al. (2004).

Although typical halos at high--{\it z} must be much less massive than their
present--day counterparts (therefore harboring fewer galaxies per halo), halos
at high--mass end may host multiple galaxies. This is consistent with the
recent discoveries that protoclusters were already forming by redshift of 4 --
6 (e.g. Ouchi et al. 2005a; Miley et al. 2004). To test for this possibility,
we have looked for close galaxy pairs that may share host halos. In light of
the luminosity segregation, we have split the full sample based on the
\wz--band magnitude because statistically selecting bright galaxies is
equivalent to selecting the sites of the most massive dark halos. Since the
spectroscopic information is not available, we count galaxy neighbors around
the brightest LBGs as a function of angular separation only.

We split the galaxies in the full ACS sample ({\it z} $\sim$ 4) into three
sub--groups, \wz $\leq 24.3$, $m^* <$ \wz $\leq 25.0$ and $m^*+0.5 \leq $ \wz
$ \leq 27.0$ where $m^*$ for the \wb--band dropouts is roughly 24.7 (M. Giavalisco et al, 
in preparation). 
The number of galaxies in each group is 29, 93 and 2150, respectively.
We consider galaxies in the two bright groups as candidates for central galaxies
in halos, and the ones in the faintest group as satellite galaxies. We count the number
of faint neighbors (\wz $\geq m^*+0.5 \sim 25.2$) around the first two bright groups as a
function of angular separation.  The upper left panel of Figure 5 shows the
average number of neighbors (cumulative counts) around these two bright groups
up to 50 arcsec with 1$\sigma$ Poisson errors, however, these errors should be
considered strictly as lower limits because these are correlated. What is
expected for an uncorrelated population with the same surface density is also
shown with the dashed line. Due to the clustering, both \wz $ \leq 24.3$ and
$24.7 <$ \wz $ \leq 25.0$ are well above the dashed line, however, \wz $
\leq 24.3$ galaxies have more neighbors than the other ($\approx 2.5\sigma$). This
cannot be understood in terms of the difference in their (large--scale) clustering strength
because the same galaxies (therefore with the same clustering strength) are 
counted as faint neighbors. In addition, there is clearly no reason to
believe that bright galaxies have more galaxies along the line of sight which
would be included as projected close--pairs. Another way to confirm that the
difference does not come from clustering is to look at the same pair counts on
larger scales.  The upper right panel of Figure 5 shows the counts at 100 -- 160 arcsec
where the two slowly converge. This is because on large scales the
contribution from halo--halo clustering dominates the counts thus reflecting
the overall clustering strength.  To remove the effect coming from the
clustering, we take the ratio of the two which is shown in the lower panels. 
The ratio exceeds unity at $\theta < 30$ arcsec and slowly goes back to unity 
on larger scales. 

This is compelling evidence that we are indeed observing central--satellite
galaxy pairs sharing the most massive halos. This is also a natural outcome of
the luminosity segregation in that the most massive halos (hosting the
brightest LBGs) are likely to have sub--structures massive enough to form
satellite galaxies bright enough for us to detect. We have tried a similar
neighborhood search for the \wu--band dropouts, but failed to detect the same
trend.  Because the \wu--band dropouts have larger area coverage thus larger
number of bright sources, we have increased the magnitude cut for the
brightest group up to $R = 22.0$ but still the brightest group does not
evidently have more neighbors than the rest. 
It is possible that the relatively shallow depth of our \wu--band data plays
an important part in the lack of faint neighbors. If, for simplicity,
no evolution of the luminosity function is assumed from $z \sim 3$ to $z\sim4$, 
one needs \wu--band dropouts fainter than $R = m^* + 0.5 = 25.0$ 
($m^* \sim 24.5$; Adelberger \& Steidel 2000) to carry out
neighbor counts at a similar luminosity level to the \wb--band dropouts case. 
In other words, we may be seeing only the bright end of the luminosity function where 
there is essentially one--to--one correspondence between host halos and galaxies. It is
consistent with Adelberger et al. (2005)'s data that have comparable depth
based on which they concluded that there is little evidence of galaxy
multiplicity and that the fraction of dark halos hosting more than one galaxy
is about 5\%. 
Another possibility, though minor, is that the relatively poor seeing of our $R$
band, which is where galaxy detection is carried out, prevents detection and
deblending of pairs that are closer than 2 arcsec.

\section{The Halo Occupation Distribution at $z\sim 4$ and $5$}

So far we have discussed the measures of the angular correlation function of
LBGs, the implied spatial correlation length, and we have reported evidence of 
a departure of the correlation function from a single power law. 
We have also shown direct evidence of sub--structure in the most massive halos 
in the form of number counts of faint LBGs around the brightest ones on scales 
$\theta \leq 10$ arcsec.
Finally, we have also measured the clustering segregation, namely the fact
that brighter LBGs are more strongly clustered in space. In order to
understand these results in light of underlying physics, we need to compare
with theoretical predictions through which quantitative interpretations of the
result can be made.

Associating the spatial distribution of galaxies with that of dark halos
requires an understanding of the relationship between galaxy properties and
those of host halos, for example, galaxy colors, stellar mass or luminosity
are correlated with circular velocity, total mass of halos and so forth. We do
not yet know if all the halos above a certain mass threshold host galaxies
with similar SEDs (colors) and luminosity to those of LBGs to be included in
our sample. At $z \sim 3$, a population of red galaxies, either heavily
obscured by dust or old, dubbed as DRGs, which are bright in the near--infrared have
been found at similar redshifts with about half the space density of \wu--band
dropouts (${\cal R} \le$ 25.5) (Franx et al. 2003; van Dokkum et al. 2003).
Many of these galaxies would not have been selected via the Lyman Break
technique.  Moreover, the submillimeter galaxies, though their space density is much
smaller (($2.4\pm1.2) \times 10^{-6}$ Mpc$^{-3}$; Chapman et al. 2003), also
coexist in the same redshift range. These high--$z$ populations (including
LBGs) suffer from respective selection biases thus cannot be solely
representative of high--{\it z} population in general.  
The halo occupation distribution (HOD) is a simple model that sets a statistical
relationship between halo mass and its probability to form galaxies, thus
provides us with a powerful tool to understand 
the characteristics of different populations independently and study their respective
relations to host halos. The HOD formalism has been successfully applied to
various local galaxy samples (e.g. different luminosities, colors, spectral
types etc.; Zehavi et al. 2004b, Magliocchetti et al. 2003) and also at
intermediate redshift (QSOs; Porciani et al. 2004).  Note that the HOD
formalism can be extended to the conditional luminosity function (CLF; Yang et
al. 2003, van den Bosch et al. 2003), to understand the relationship between
halo mass and galaxy luminosity. We are currently working to constrain this
relationship at $z\sim 3, 4$ and $5$ using very large samples of LBGs
extracted from the COSMOS 2--square degree survey,
which provide excellent sampling at the high end of the
luminosity (and hence mass) distribution (K.-S. Lee et al. in preparation).

We move on now to consider a simple model for the halo occupation
distribution function, which can be used to understand the
phenomenology of the observed clustering properties of LBGs in physical terms,
and derive its parameters. These include the minimum mass of halos that can
host the observed galaxies and the average halo mass for each galaxy
sub--sample. We will then use the best--fit HOD parameters to discuss the inferred
halo bias and its implications.

\subsection{The HOD Formalism and its Physical Implications}

The angular correlation function can be considered to
have two separate contributions, galaxy pairs from the same halo and from two
distinct halos.
\begin{equation}
w(\theta) = w_{1h}(\theta) + w_{2h}(\theta)
\end{equation}
On large scales, only $w_{2h}(\theta)$ contributes to the total angular
correlation function because any galaxy pair with a large angular separation
cannot reside in the same halo. On the other hand, on small scales and
on scales comparable to virial diameter of halos, both terms will be present
with a ratio that depends on the degree of galaxy multiplicity. The two--halo
term can be constrained from clustering measurements; however, the galaxy
multiplicity function or halo occupation distribution (HOD) is not very well
constrained at high redshifts because of the still relatively small and shallow
sample. Bullock et al. (2002) estimated the HOD from a sample of 800 LBGs with
spectroscopic redshifts at {\it z} $\sim$ 3 based on a ground--based survey
(${\cal R} \le$ 25.5; Steidel et al. 1998). Although this sample has the
advantage of having spectroscopic information, the median luminosity of the
sample is significantly brighter than our ACS sample at $z\sim 4$, and thus
the number of faint members of pairs or multiplets is very likely quite
small. Adelberger et al. (2005) estimate that at most 5\% of their LBGs at $z
\sim 3$ (an extended version of the sample used by Bullock et al. 2002) are
members of close pairs. We do not yet know the luminosity function of
satellite LBGs; however, our neighborhood analysis discussed earlier for the
\wb--band dropout sample at $z\sim 4$ showed that a large fraction of the excess
number of neighbors comes from \wz $>26$ (galaxies counted as faint neighbors
have $25.2\leq$\wz$\leq27.0$).  Using our template LBG spectrum, \wz = 26.0
corresponds to ${\cal R}=25.5$ at $z \sim 3$, that is the limiting flux of
Adelberger et al.'s sample. While the bright members of pairs and multiplets
are certainly present in their sample, Adelberger et al.'s sample must be
highly incomplete for satellite galaxies.

Once the halo power spectrum is given and an HOD model assigned, one can
unambiguously determine the correlation function (Seljak 2000; Berlind \&
Weinberg 2002; Berlind et al. 2003) with suitable assumptions as to how
galaxies are distributed within halos (see later). We adopt a simple functional
form (Jing et al. 1998; Wechsler et al. 2001).  This model assumes that all
galaxies are associated with dark matter halos and makes a simplifying
assumption that the number of galaxies within a given halo depends only on the
halo mass and can be modelled by a single power--law:
\begin{eqnarray}
\langle N_g(M) \rangle & = & (M/M_1)^\alpha \; \; \; \; if\; M \geq M_{min} \nonumber \\
&=& 0 \; \; \; \; \; \; \; \; \; \; \; \; \; \; \; \; \; \; \; \; otherwise
\end{eqnarray}
where $M_1$ is the typical halo mass at which there is on average one galaxy
per halo, $M_{min}$ is the cutoff halo mass below which halos cannot host
galaxies and $\alpha$ is the power--law index, the efficiency of galaxy
formation within halos of given mass {\it M}.

While these three parameters, $M_1,M_{min}$ and $\alpha$ are sufficient to
constrain the correlation function on large scales, the shape of the
correlation function on small scales also depends on how galaxies populate
host dark halos. This is because the one--halo term $w_{1h}$ is
proportional to $\langle N_g (N_g - 1) \rangle$ while $w_{2h}$ is proportional to $\langle
N_g \rangle$ (see Appendix A. for details).  Berlind \& Weinberg (2002)
discussed what these effects are and their impact on the shape of the
correlation function on small scales. This includes the scatter of $N_g$
around the mean, galaxy concentration with respect to halo concentration, the
existence of a central galaxy at all times and the galaxy velocity dispersion
with respect to that of halos. These factors can be used to construct more
sophisticated models to constrain the HOD once a larger galaxy sample becomes
available.  However, at high redshift, the number of galaxies included in any
sample (so far) remains much smaller than that in such surveys as 2dF or SDSS,
hence a simple approach seems preferable.  Therefore we make simplifying
assumptions to this model as follows. We force the first galaxy to be placed
at the center of host halos and the rest (satellite galaxies) largely follow
the halo mass distribution for which we adopt an NFW profile (Navarro, Frenk
\& White 1997).  The latter is consistent with the finding of Kravtsov et
al. (2004) that the sub--halos in their simulations follow the distribution of
dark matter within the halos.

In order to calculate \wth\, we closely follow the recipe given by
Hamana et al.  (2004), using the procedure described in Appendix A. We
generate a library of model \wth's for a wide range of ($M_1,\alpha$)
and determine the best--fit \wth\ that matches the abundance ($n_g$) and
the clustering strength ($M_{min}$, $\alpha$ and $M_1$) simultaneously using
$\chi^2$ minimization. Note that $M_{min}$ is not a free parameter once the
space density is specified.
\begin{eqnarray}
n_g (z) = \int_{M_{min}}^{\infty} n_h(M) \langle N_g(M)\rangle dM \nonumber\\
\langle n_g \rangle = \frac{\int dz N(z) [dV(z)/dz] n_g(z)}{\int dz N(z) [dV(z)/dz]}
\end{eqnarray}
where $N(z)$ is the redshift selection function obtained from the simulations and $n_h(M)$
is the halo mass function for which we adopt the analytic halo model by 
Sheth \& Tormen (1999, 2002).
Note that we derive the HOD parameters, $M_1$ and $\alpha$, from the \wth\ of the 
full sample (B270 and V270) and do not attempt to constrain these parameters separately
 for the sub--samples.  We also fix the observed number density to the incompleteness--corrected
number density integrated down to \wz = 27.0 for the fit. 

Table 4 lists the HOD parameters obtained using this method for the \wb\ and
\wv--band dropouts only. For the \wu--band dropouts, we are unable to put any
robust constraints on these parameters primarily because the HOD fit is most
sensitive to \wth\ at small separations.  The \wth\ measure of the \wu--band
dropouts on such scales seems to indicate a constant value rather
than steepening. 
This is more consistent with the correlation function having the
two--halo term only, i.e. each observed galaxy belongs to a different halo,
in which case $w_{2h}$ stays relatively constant on small scales ($\xi(r)$
actually falls off but the projection, \wth, stays constant due to the
contribution from projected galaxy pairs). Figure 6 shows the best--fit model
\wth\ for the full sample of \wb\ and \wv--band dropouts with the
observational measures.  The corresponding confidence levels for the HOD fits
are illustrated in Figure 7.

The best--fit parameters for the full samples are ($M_{min}$, $\alpha$, $M_1$) =
($7\times10^{10} h^{-1}M_{\sun}$, 0.65, 1.3$\times 10^{12} h^{-1}M_{\sun}$) for the \wb--band
dropouts and ($5\times10^{10} h^{-1}M_{\sun}$, 0.80, $1.0\times 10^{12} h^{-1}M_{\sun}$) for
the \wv--band dropouts, respectively. Though galaxies were selected with different filter
systems and also their sample is shallower, it is interesting to compare them to those of
Hamana et al. (2004), ($1.6\times10^{11} h^{-1}M_{\sun}$, 0.5, $2.4\times
10^{12} h^{-1}M_{\sun}$), ($1.4\times10^{11} h^{-1}M_{\sun}$, 0.5, $1.4\times
10^{12} h^{-1}M_{\sun}$) for their LBG$z$4s and LBG$z$5s samples. $M_{min}$ is found
to be a factor of 2 -- 3 smaller from our survey. This is hardly surprising
because $M_{min}$ depends mainly on the depth of given survey.  Our sample is
significantly deeper than theirs ($\approx 1$ mag) and, because LBGs have
compact morphology, more complete in the magnitude range where there is overlap. As a
result, the overall clustering strength of our sample is weaker which is most
directly reflected in $M_{min}$. The parameter $M_1$, the mass scale at which
statistically there is one galaxy corresponding to each halo, is smaller in
our case, consistent with our expectations.  
On the other hand, this work and 
previous studies consistently show that $\alpha$ should be 0.5 --1 though it may 
be dependent on $M$ (thus the depth). Therefore over
a reasonable range of observable halo mass it can be approximated (to the zeroth
order) as a constant, thus nearly independent of mass scale. Unfortunately, as
can be seen in Figure 7, it is the least constrained parameter from the HOD fit. In
principle, however, $\alpha$ can be constrained independently via studies such
as our neighborhood analysis or gravitational lensing surveys.

We calculate $\langle N_g \rangle_M$ (note the subscript ``$M$'' to
distinguish the quantity from the previously mentioned $\langle N_g(M)
\rangle$), 
the ratio of galaxy number density to halo number density (=$n_g/n_h$) 
integrated over the allowed range of halo mass. 
We also calculate $\langle M \rangle$, the average mass of halos that host the observed LBGs
(assuming the best-- fit HOD parameters) as follows:
\begin{eqnarray}
\langle N_g \rangle_M &=& \frac{\int_{M_{min}}^\infty n_h(M)\langle N_g(M)\rangle dM}{ \int_{M_{min}}^\infty n_h(M) dM}  \nonumber \\
\langle M \rangle &=& \frac{\int_{M_{min}}^\infty M \langle N_g(M) \rangle n_h(M) dM}{\int_{M_{min}}^\infty \langle N_g(M) \rangle n_h(M)dM}
\end{eqnarray}

For each sub--sample, we compute $M_{min}$ that matches the number density,
then $\langle N_g \rangle_M$ and $\langle M \rangle$ are computed using the
best--fit HOD parameters. Table 4 summarizes the results. Uncertainties are
computed by propagating the bootstrap errors of \wth\ after marginalizing over
$M_1$ and $\alpha$.  
We find $\langle N_g \rangle_M = 0.3$ and $0.2$ for the full sample of
\wb--band and \wv--band dropouts, respectively. If each galaxy belongs to 
a different halo, this quantity would represent the average number of galaxies
per halo. For example, for our \wb--band dropouts, this means that only 30\% 
of the halos harbor LBGs bright enough to be detected 
(in other words, 30\% of the halos that are more massive than $M_{min}$).
In reality, where multiple galaxies can share host halos, $\langle N_g \rangle_M$
can be regarded as an upper limit.
Note that Adelberger et al. (2005) and Giavalisco \& Dickinson (2001), based on the
ground--based sample with ${\cal R} \leq 25.5$, argued that most halos at {\it
z} $\sim$ 3 host visible LBGs. While it is not clear if their claim is in
quantitative discrepancy with our estimate, we note that possible evolutionary
effects from $z\sim 4$ to $z\sim 3$ remain poorly understood at this time. It
is interesting, however, that the \wb--band dropouts with \wz $ \leq 26.0$
(B260), which have a comparable absolute luminosity threshold to that of the
\wu--band dropouts of ${\cal R} \leq 25.5$ (U255), have $\langle N_g \rangle_M
\sim 0.5$, similar to what Adelberger et al. (2005) have found (see their Figure 10).
This suggests that LBGs at $z \sim$ 3 and 4 of similar luminosity
also have a similar duration of star--formation during which they are visible 
(duty--cycle hereafter). 
Note that if the average galaxy luminosity is closely tied to the halo mass,
as suggested by the clustering segregation, then in principle one can
empirically find the intrinsic relationship between UV luminosity and mass at
a given redshift\footnote{This would be the average ``observed'' UV
luminosity, not corrected for the effects of dust obscuration. Since
not all galaxies have the same dust obscuration, and since the amount of dust
obscuration is, in general, dependent on the UV luminosity itself, the
statistical relationship between obscured and unobscured UV luminosity and
mass will be, in general, different.}. For example, given a mass spectrum and a
HOD function, such a relationship will simulataneously reproduce the shape of
the correlation function, that of the clustering segregation, and the
luminosity function.
Interestingly, when the same absolute
magnitude cut is made to the \wv--band dropouts, we find that $\langle N_g
\rangle_M$ is a factor of 2 -- 3 smaller. We postpone the interpretations of these results
to the next subsection.  
Finally, we note explicitly that the clustering properties of LBGs are consistent
with the notion, built in the HOD model itself, that $\langle N_g \rangle_M$
becomes smaller for less massive halos. A possible physical explanation for this 
fact could be that in low--mass halos, gas cooling (hence star formation)
is much more inefficient due to photo--ionization (squelching) or SNe feedback
(e.g. Silk 1997, Efstathiou 2000). In addition, it is
possible that the duty--cycle of galaxies may be mass--dependent such that
lower--mass halos have shorter duty--cycle, thus resulting in a smaller value
of $\langle N_g \rangle_M$.

\subsection{A Different Approach: Separation of Central and Satellite Galaxies}

We have also explored an alternative approach, which is to model central
galaxies and satellites separately (Guzik \& Seljak 2002; Kravtsov et al. 2004; Zheng et al. 2005)
In this model, the probability for a halo to have a central 
galaxy is a step function, namely either the halo has a central galaxy or it does
not,  whereas the distribution of satellite galaxies follows a
power--law:  
\begin{eqnarray}
\langle N_g (M) \rangle &=& 1 + (M/M_1')^{\alpha'} \; \; \; \; if \; M \geq M'_{min}\nonumber  \\ 
&=& 0 \; \; \; \; \; \; \;\;\;\;\;\; \;\; \; \; \; \; \; \; \; \; \; \; \; \; \;otherwise
\end{eqnarray}
We refer to this model as the ``one--plus'' model as opposed to the previous one
which we refer to as the ``pure power--law'' model hereafter.  Note that the
normalization $M_1'$ and the threshold mass $M'_ {min}$ have different
physical meanings in this model compared to the previous one. In the present
model the parameter $M_1'$ is the halo mass at which there is one satellite
galaxy (thus two galaxies in total) in the host halo. The parameter $M'_{min}$
is conceptually the same as $M_{min}$ of the first model, except that now
halos above this threshold value always have $\langle N_g \rangle \geq 1$,
while in the pure power--law model only a fraction of halos with $M>M_1$ are
host to more than one galaxy on average.
Using the alternative model for the halo distribution function of Lyman--break
galaxies we find that the best--fit parameters for the \wb--band dropouts are
($M'_{min}, \alpha', M_1') = (2\times10^{11} h^{-1}M_\sun, 1.3, 4 \times10^{12}
h^{-1}M_\sun$).

We have fit the two HOD models to our data, and the $\chi^2$ analysis shows
that both reproduce the observed \wth\ equally well. Figure 8 shows the
predicted \wth\ from the best--fit HOD parameters in both cases together with
the observed correlation function of the \wb--band dropouts. We found
$\chi^2_{1+}= 0.97$ and $\chi^2_{ppl}=0.91$. 
 In addition, they predict similar $\langle N_g \rangle$ for mass scales of
$10^ {12}$ -- $10^{13}$ $h^{-1}M_\sun$, to which most of massive halos
included in the sample belong. This is shown in the right panel of Figure
8. At the high--mass end ($M > 10^{13} h^{-1}M_\sun$), however, the two models
diverge from each other rather quickly. To constrain the HOD function in this
mass range one needs a much larger sample covering significantly more area
than the GOODS, because of the rarity of high--mass halos. On scales
$M<10^{12} h^{-1}M_\sun$, on the other hand, the ``pure power--law'' model
predicts numerous low--mass halos being included in the sample with a
relatively low efficiency of actually hosting a galaxy, while for the
``one--plus'' model, $\langle N_g \rangle$ is truncated (by definition) at the
mass scale of $M < 2\times10^{11}h^{-1}M_\sun$.

Finally, we observe that although the two models are equally good at
reproducing the observations, this is very likely the result of the still
relatively large uncertainties in the measure of \wth.  The rationale behind
the ``one--plus'' model is that every single halo above $M_{min}$ participates
in forming observed galaxies. In contrast, the ``pure power--law'' model
allows a fraction of halos to form galaxies over some range of mass ($M <
M_1$), with the complementary fraction remaining quiscent (either because no
star formation has not started in them or becuase observed in between bursts
of star formation). The currently available data at high rdshifts do not allow
us to test which model is a better representation of reality. 
However, we note that the built--in assumption of
the ``one--plus'' model, that there is always one central galaxy (in our case,
one LBG), in every halo above a mass threshold, naturally precludes the
possibility of having other types of galaxies (i.e. non--LBG) as the central
galaxy in those halos. While such a limitation is not a problem in the local
universe, it is very likely a misrepresentation of the situation in the
high--redshift universe. For example, a number of observations have shown the
existence of massive galaxies at $z\sim 3$ whose UV spectral energy
distribution is such that they would not be selected as Lyman-break
galaxies. Such galaxies very likely have masses that are in the high--mass end
of the LBG mass range (e.g. Shapley et al. 2004, van Dokkum et al. 2004).  The
``one--plus'' model would not provide a fair representation of the HOD if such
galaxies are relatively common at high redshift.  For these reasons we will
limit the discussion only to the ``pure power--law'' model in what follows.

Realistically, we expect observed LBGs to be the central galaxies in only a 
certain fraction of halos ($M \geq M_1$). For this reason, we expect that
$\langle N_g \rangle = p + (M/M_1)^\alpha$ ($M>M_{min}$ with $0<p<1$) may be a
more reasonable description of the actual halo occupation distribution such
that a fraction $1-p$ is reserved for other populations. While $M_1$ and
$\alpha$ (to a lesser extent) may depend on the depth of a given survey, $p$
more likely represents a fundamental quantity that is the efficiency of LBGs
as central galaxies on the high end of the halo mass function.
We plan to explore this possibility in a follow--up paper with a larger
data set (K.-S. Lee et al., in preparation). 

\subsection{The Evolution of Halo Bias with Redshift}

We compute the nonlinear autocorrelation function of matter $\xi_m$(r) at
redshift {\it z} $\sim$ 3, 4 and 5 by adopting the algorithm by Peacock \&
Dodds (1996). $\xi_m$(r) is then inverted to $w_m(\theta)$ using the Limber
transform as follows:
\begin{equation}
w_m(\theta) = \frac{\int_0^\infty dz N^2(z) \int_{-\infty}^{\infty}[dx/R_H(z)]\xi_m([D_M(z)\theta]^2+x^2)^{1/2}}{[\int dz N(z)]^2}
\end{equation}
where $D_M(z)$ is the proper motion distance and $R_H(z)$ the Hubble radius at
given redshift {\it z}. Using this method, the average bias is calculated as a
function of angular separations.

\begin{equation}
\langle b_{obs}(\theta) \rangle= \sqrt{\frac{w(\theta)}{w_m(\theta)}}
\end{equation}

The bias $\langle b_{obs}(\theta) \rangle$ monotonically decreases at small separations
then stays constant at $\theta \geq 70$ arcsec. We adopt the value of $\langle
b_{obs}\rangle$ at $100$ arcsec as our linear bias value.  The theoretical
predictions of the galaxy and halo bias can be computed from the given halo
model as follows:
\begin{eqnarray}
\langle b_g \rangle &=& \frac{1}{n_g}\int_{M_{min}}^{\infty} n_h(M)b_h(M)\langle N_g(M) \rangle dM \nonumber \\
\langle b_h \rangle &=& \frac{1}{n_h}\int_{M_{min}}^{\infty} n_h(M)b_h(M)dM 
\end{eqnarray}
where $n_h = \int_{M_{min}}^{\infty} n_h(M)dM$ and 
$n_g = \int_{M_{min}}^{\infty} n_h(M)\langle N_g(M)\rangle dM$.
We use the best--fit HOD parameters for $\langle N_g(M)\rangle$.  Note that halo
bias $\langle b_h \rangle$ is HOD--independent whereas $\langle b_g \rangle$
is not.  The observed galaxy number density is calculated by fitting the full
sample to the Schechter luminosity function after correcting for photometric
incompleteness (e.g. Giavalisco 2005) and integrating down to each magnitude
cut (see Table 3).  The error bars in the number density are derived by
propagating the uncertainties from the Schechter LF fit.

Figure 9 shows the average galaxy bias from the clustering measures together
with the predictions of $\langle b_h \rangle$ and $\langle b_g \rangle$ at
{\it z} $\sim$ 4 and 5. Note that the abscissa represents the halo abundance
($n_h$) for the dashed lines and the galaxy abundance ($n_g$) for the solid
lines and symbols ($n_g = n_h$ if $N_g \equiv 1$).  The observed galaxy bias
values are in good agreement with the CDM model predictions when the best--fit
HOD parameters are assumed.

To compare the results from different cosmic epochs, it is important to remove
the effect of the luminosity segregation which could be mistakenly interpreted
as ``evolution''.  We truncate each sample to the same absolute luminosity
threshold to match that of our brightest sample, the \wu--band dropouts ($R$
$\leq$ 25.5), the absolute magnitude of which is $M_{1700} \leq$ -20.0. This corresponds
to $R$ $\leq$ 25.5, \wz $ \leq$ 26.0 and \wz $\leq$ 26.6 at {\it z} $\sim$ 3, 4 and 5,
respectively.  Since these luminosity thresholds are similar to those of the
U255, B260 and V265 sample, we compare the bias values for these magnitude
cuts. Figure 10 illustrates how these bias values change with redshift compared
to halo biases for various mass thresholds.  The dashed lines indicate the
evolution of the average halo bias with a constant halo mass threshold for $M_{halo}$
= $10^{10}, 5\times10^{10}, 10^{11}, 5\times10^{11}, 10^{12}$ and $5\times
10^{12}$ $h^{-1}M_\sun$ (from bottom) from the Sheth \& Tormen (1999) model.  Lines
serve as as a guide to determine the effective halo mass threshold of a given
galaxy sample.

It is intriguing that our measures at {\it z} $\sim$ 3 and 4 imply that these
galaxies are hosted by halos of mass $5\times10^{11} h^{-1}M_\sun \lesssim M_{halo}
\lesssim 10^{12} h^{-1}M_\sun$, however, at {\it z} $\sim$ 5, the host halo mass is
approximately $10^{11}h^{-1}M_\sun$, a factor of 5 -- 10 times smaller than its
lower--{\it z} counterparts at a given fixed luminosity. Hamana et
al. (2004) reported the average bias of their LBG$z$4s with similar absolute
luminosity threshold, ${\it i}^\prime \leq 26.0$, to be 3 -- 4.5, consistent
with our results.  This implies that at $z \sim 5$, star--formation
may have been more efficient, hence galaxies of comparable luminosities were
hosted in much lower--mass halos. 
This result is, in fact, another manifestation of what we
have mentioned in an earlier subsection, that $\langle N_g \rangle$ was a factor
of 2 -- 3 smaller at higher redshift when the same absolute
luminosity cut was made for a given choice of HOD parameters.

\section{Summary}

1. We have studied the spatial clustering properties of LBGs at three
different cosmic epochs.  We used three samples of LBGs which include two deep
ACS--selected LBGs samples at $z \sim$ 4 and 5 down to a magnitude limit of
\wz $ \sim 27$ and the ground--based sample at $z \sim 3$ down to $R$ $\sim$
25.5. The ACS samples enable us to study the spatial distribution of the
large--scale structure as well as of small--scale clustering whereas the
ground--based sample allows us to probe the large--scale clustering of bright LBGs
selected from a large contiguous area.

2. We found that the clustering strength (the amplitude of \wth\ or value of
$r_0$) of LBGs at all redshifts that we have studied depends on the UV
luminosity (rest--frame $\lambda$ $\approx$ 1700 \AA) of the galaxies. This luminosity
segregation is in good quantitative agreement with previous investigations
based on much brighter galaxy samples and we are able to robustly extend this
reesult down to at least 1 mag fainter than previous
studies. The faintest galaxy sub--sample has a real--space correlation length as
small as 2 $h^{-1}$ Mpc, only half of that known for ${\cal R} \leq 25.5$
sample. A physical implication is that galaxies brighter at UV wavelengths are
hosted by more massive dark halos, implying that the star--formation is
primarily regulated by local gravity, and suggesting that other physical
mechanisms (major mergers, interactions) are secondary drivers.

3. We have carried out neighbor counts around bright LBGs in our sample
and counted the number of bright--faint galaxy pairs. We discovered that the
likelihood of finding faint neighbor galaxies (\wz $\gtrsim m^* + 0.5$) 
around bright
ones (\wz $ \leq 24.3$) up to scales comparable to the virial radius of dark
halos, is at least 20 -- 30 \% higher than random LBG--LBG pairs with a high
significance. 
This is consistent with the interpretation that there are faint companions around
the brightest galaxies sharing host halos, supporting that the departure of the 
correlation function on small scales (from the extrapolation of the large--scale 
power--law) is indeed due to central--satellite pairs.
The observed sub--structure is also consistent with the luminosity segregation
that we have detected, in that brighter galaxies (more
massive halos) are likely to have fainter companions (halo sub--structures).
We do not yet know the relationship between the UV luminosity and the actual
host halo mass. However, the clustering segregation
gives support to the interpretation that the most luminous LBGs are hosted by the
most massive halos and tend to have more satellite galaxies in halo sub--structures. 
Due to the relatively small dynamic range in magnitude of our survey, we are
limited to a bright cut of \wz = 26.0 and also to cumulative magnitude
thresholds for our samples, however, we expect this effect to be more dramatic
in future surveys that cover larger cosmic volume, which should confirm our results.

4. Our measurement of the correlation function at $z \sim$ $4$ and $5$ indicates
a steepening on small scales, thus cannot be well described by a simple power--law.
We also find that the scale at which the steepening
occurs is comparable to the angular size of dark halos at the relevant redshift
range, and hence we attribute this steepening to the presence of multiple 
galaxies within the same halo.  A simple HOD model with a suitable scaling law
between halo mass and the number of galaxies seems to be able to explain both
the large--scale and small--scale behavior of the angular correlation function. We
find the best--fit HOD parameters for LBGs at {\it z} $\sim$ 4 and 5 to be ($M_{min}$
, $\alpha$, $M_1$) = ($7\times 10^{10} h^{-1}M_\sun$, 0.65, $1.3 \times 10^{12} h^{-1}M_\sun$)
and ($5 \times 10^{10} h^{-1}M_\sun$, 0.8, $1.0 \times 10^{12} h^{-1}M_\sun$), respectively. With
these parameters, we find that, on average, the number density of galaxies is
a factor of 4 -- 5 smaller than that of dark halos for the full sample and that
this discrepancy is smaller for brighter galaxy samples. This suggests that
the duty--cycle of faint LBG populations may be shorter than that of 
brighter counterparts suggested by shallower surveys, but at the same 
absolute luminosity cut, our results are in good agreement with shallower
surveys.  However, at {\it z} $\sim$ 3, our results do not indicate a
similar steepening, but is more consistent with a constant value for the
correlation function at small scales.  We interpret this discrepancy to be caused by
the difference in survey depth, as shallower surveys tend to pick up only bright
LBGs which probably tend to be central galaxies
and not satellites (which are usually fainter) within halo sub--structures that
largely produce this small--scale feature.

5. The scaling of the correlation length or galaxy bias as a function of
volume density shows a similar trend to that expected for dark halos, implying
that our measures fully support the biased galaxy formation scenario predicted
by the cold dark matter framework.  This suggests that LBGs flag the sites of
dark halos efficiently, however, not every halo is lit up at the same
time. This has to do with the fact that the duty--cycle of LBGs is shorter
than the cosmic time spanned by the observations (0.3 -- 0.5 Gyr), therefore
only a fraction of halos are lit up by star--formation at any given time.  In
addition, a HOD that scales with halo mass might be an indication that
duty--cycle is also mass--dependent such that higher--mass halos may have
longer duty--cycles than their lower--mass counterparts.

6. We truncate each sample to a fixed absolute magnitude and compute the average
bias of these galaxies. 
At $z \sim 5$, galaxies with a given luminosity are hosted by halos of mass $\sim$
$10^{11} h^{-1}M_\sun$ whereas at $z \sim 3$ and $z \sim 4$, the corresponding host halos are 
a factor of 5 -- 10 more massive. The implication is that at $z \sim 5$
star--formation was more efficient than the later epochs.

\acknowledgments
We thank Cristiano Porciani for very helpful assistance. KL acknowledges 
Takashi Hamana for kindly providing his halo code for model comparisons. 
MG and KL thank David Weinberg for his very useful comments on HOD modelling.
We would also like to thank an anynymous referee for useful suggestions. 
We are grateful to the entire GOODS team and particularly, Tomas Dahlen,
who has put great efforts into calibrating many of the GOODS ground--based 
data sets, which made our ground--based measures possible.

\appendix

\section{\wth\ and Halo Occupation Distribution}

For a given set of $(M_1,\alpha)$ with a fixed number density $\langle n_g \rangle$, the
minimum mass $M_{min}$ is determined by matching the number density of the
observed galaxies $\langle n_g \rangle$ with that expected from the halo model as
follows:
\begin{equation}
n_g(z) = \int_{M_{min}}^\infty dM n_h(M)\langle N_g(M)\rangle
\end{equation}
\begin{equation}
\langle n_g \rangle = \frac{\int dz N(z)[dV/dz]n_g(z)}{\int dz N(z)[dV/dz]}
\end{equation}
where $N(z)$ is the normalized redshift selection function and $dV/dz$ is the
comoving volume element per unit solid angle.  The contributions from the one--halo
and two--halo components are computed from their respective galaxy power
spectra, $P^{1h}_g(k,z)$ and $P^{2h}_g(k,z)$. \wth\ is the inverse Fourier
transform of the total power spectrum, $P_g(k,z)$:
\begin{equation}
w(\theta) = \int dz N(z)^2 (\frac{dr}{dz})^{-1} \int dk \frac{k}{2\pi} P_g(k,z)J_0(r(z)\theta k)
\end{equation}
where $J_0$ is the Bessel function of the first kind and $r(z)$ is the radial comoving distance. \\ The one--halo
contribution to the power spectrum is given as:
\begin{equation}
P_g^{1h}(k,z) = \frac{1}{n_g(z)^2}\int dM \frac{dn_h}{dM}\langle N_g (N_g - 1)\rangle \vert y(k,M)\vert^p
\end{equation}
where $y$ is the normalized Fourier transform of the halo mass profile (Seljak
2000; Scherrer \& Bertschinger 1991). We further assume that each halo has an
NFW profile with the concentration parameter varying with the halo mass as
found by Bullock et al.  (2001).  The parameter $p$ is taken to be $p=1$ if $
\langle N_g (N_g -1) \rangle < 1$, and $p=2$ otherwise (Seljak 2000).

The two--halo term contribution to the galaxy power spectrum according to the 
linear halo bias model (Cole \& Kaiser 1989; Mo \& White 1996) is:
\begin{equation}
P_g^{2h} (k,z)=P_{lin}(k)\{\frac{1}{n_g(z)}\int dM \frac{dn_h}{dM}\langle N_g(M)
\rangle b(M)y(k,M) \}^2
\end{equation}
where $P_{lin}(k)$ is the linear dark matter power spectrum and $b(M)$ is the 
linear halo bias from the fitting function of Sheth \& Tormen (1999).


\clearpage

\begin{figure}
\epsscale{.90}
\plotone{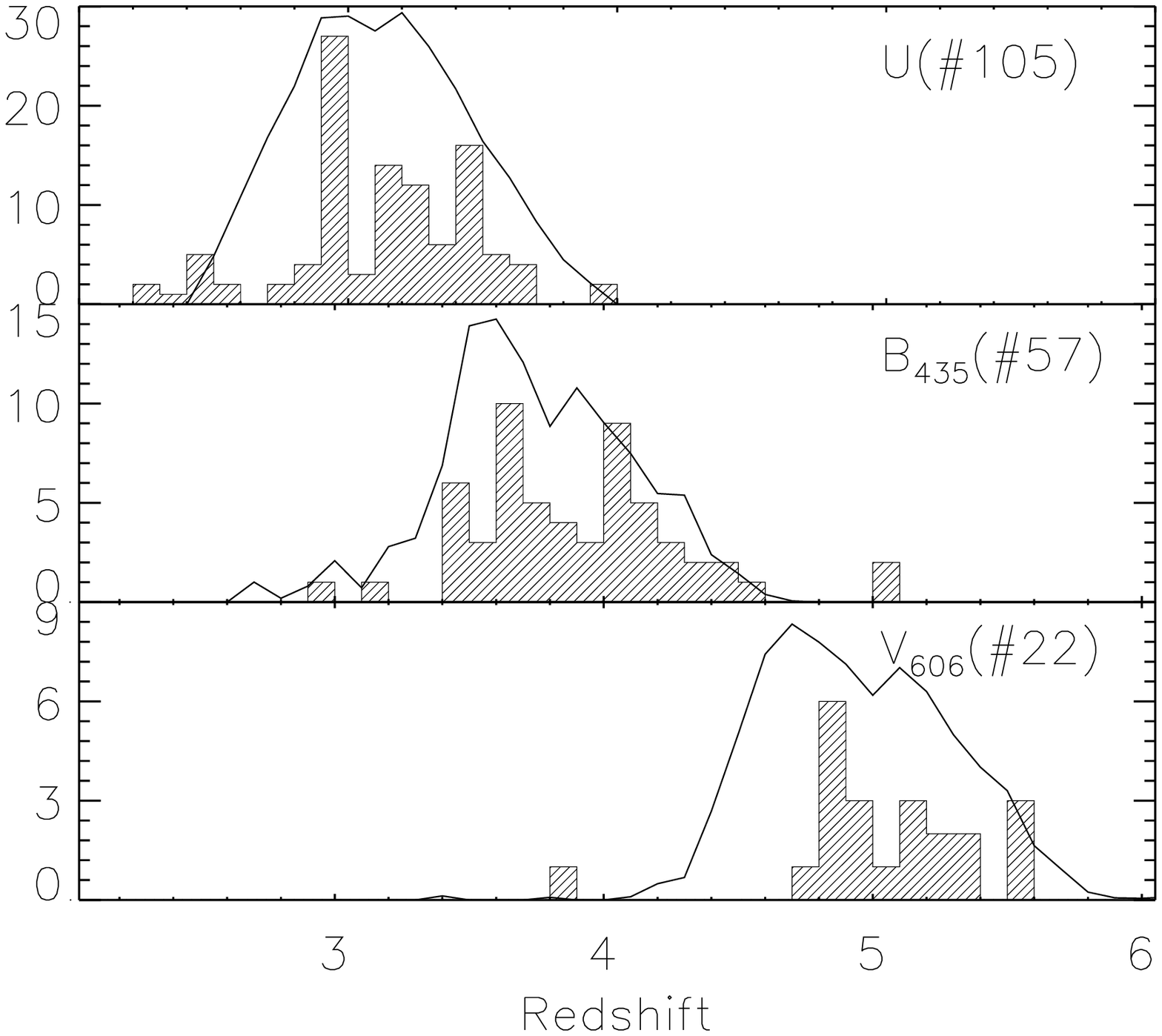}
\caption{The redshift distribution functions $N(z)$ of 
the \wu, \wb\ and \wv--band dropouts (from top) estimated from Monte Carlo simulations. The coarser
histograms indicate the distribution of spectroscopically confirmed objects
to date whose numbers are specified on top right corner of each panel. 
The histograms from the simulations are
arbitrarily normalized to match the distribution of the spectroscopic
samples.}
\end{figure}

\clearpage

\begin{figure}
\epsscale{1.00}
\plottwo{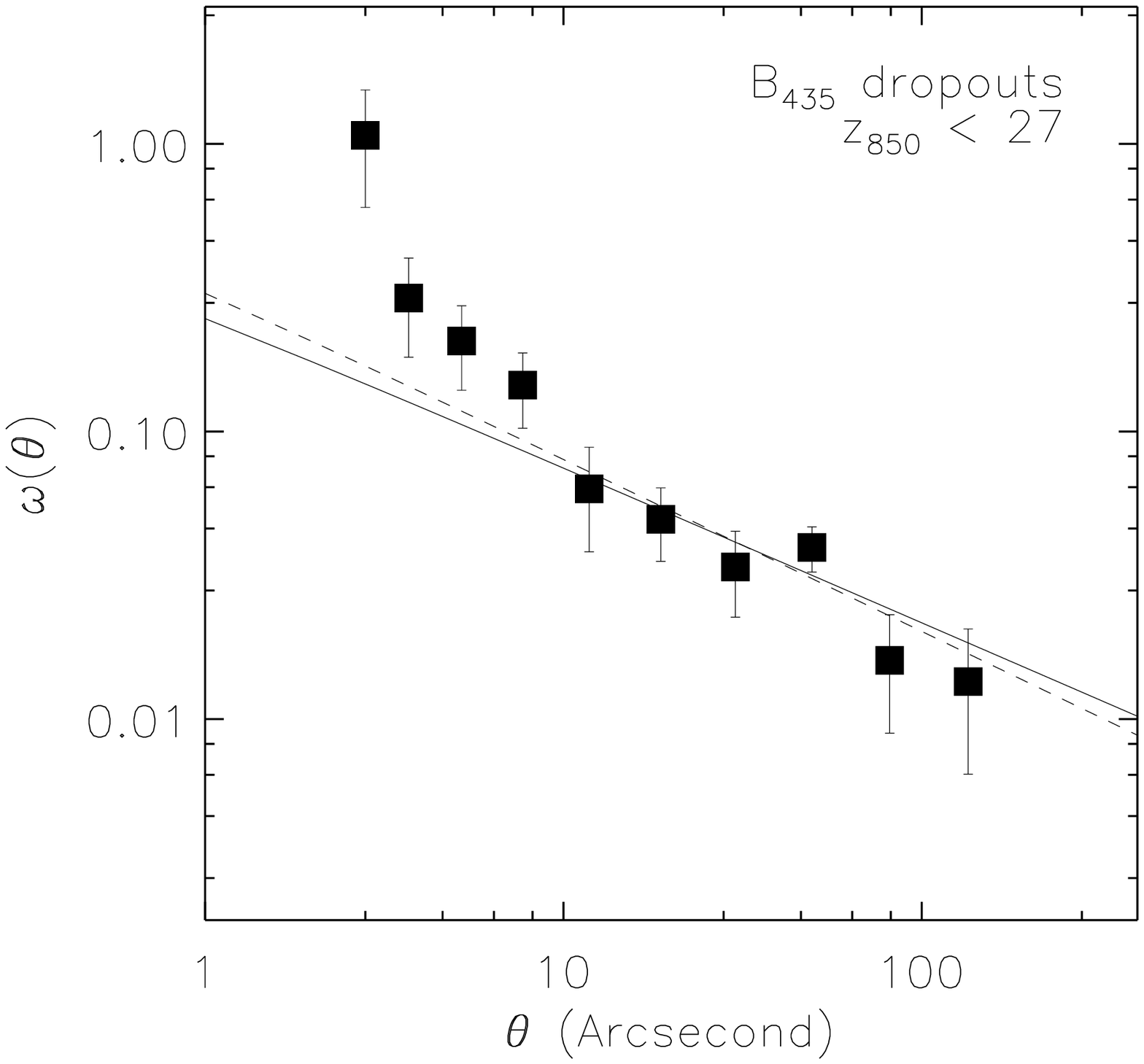}{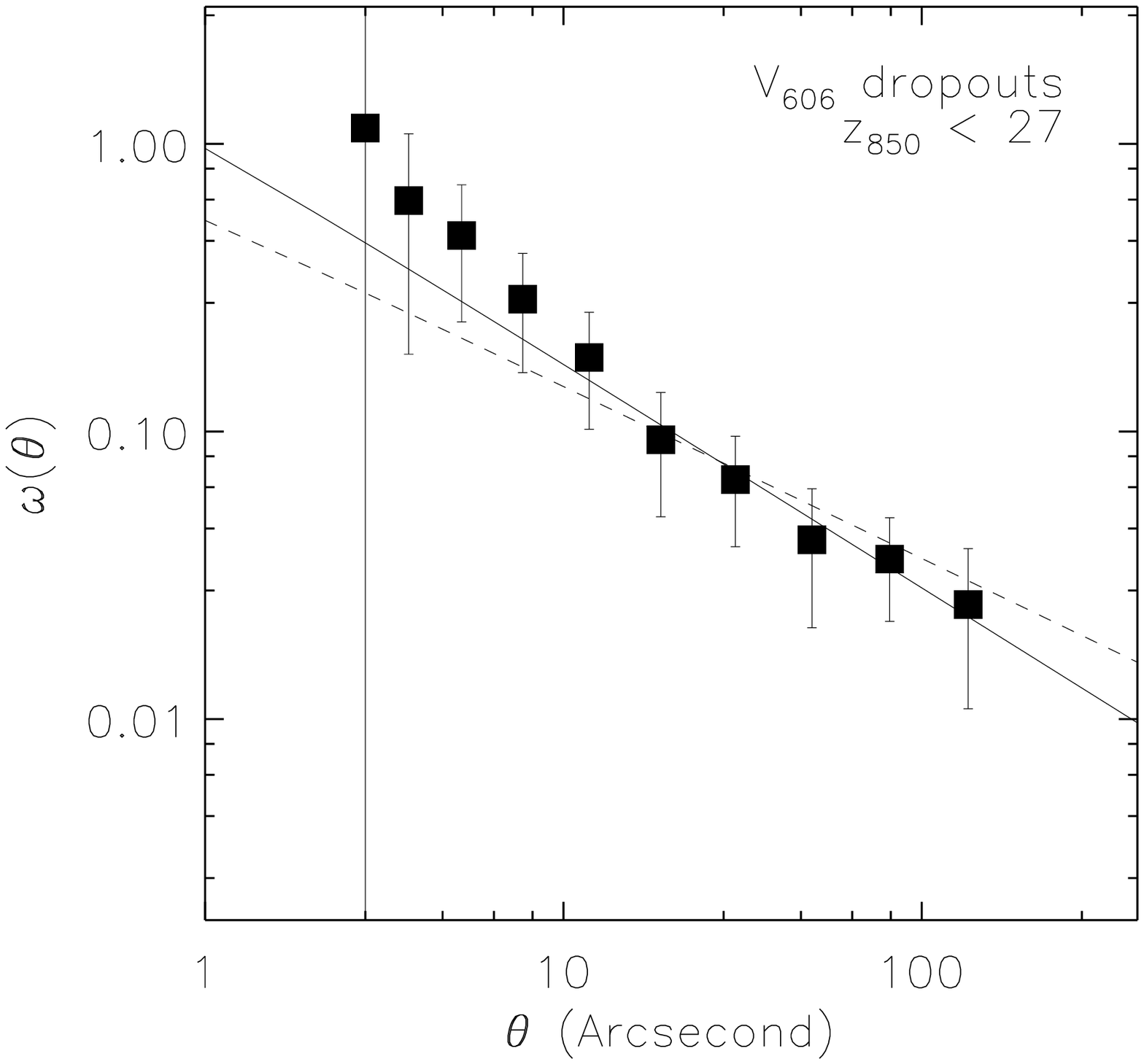}
\caption{The angular correlation function \wth\ of the \wb\
 (the left panel) and \wv--band dropouts (the right panel). The points are the 
measurements corrected for the integral constraint ($IC$)
together with the best--fit power--laws. The solid line shows the results when the
slope is allowed to vary and the dashed line shows the result when the slope is fixed to 0.6.
The fit is done including angular separations $\theta >$ 10 arcsec only. While
the data can be well described by a single power--law on large scales, a significant
departure from a power--law on scales $\theta < 10$ arcsec is observed.}
\end{figure}

\clearpage

\begin{figure}
\epsscale{1.00}
\plotone{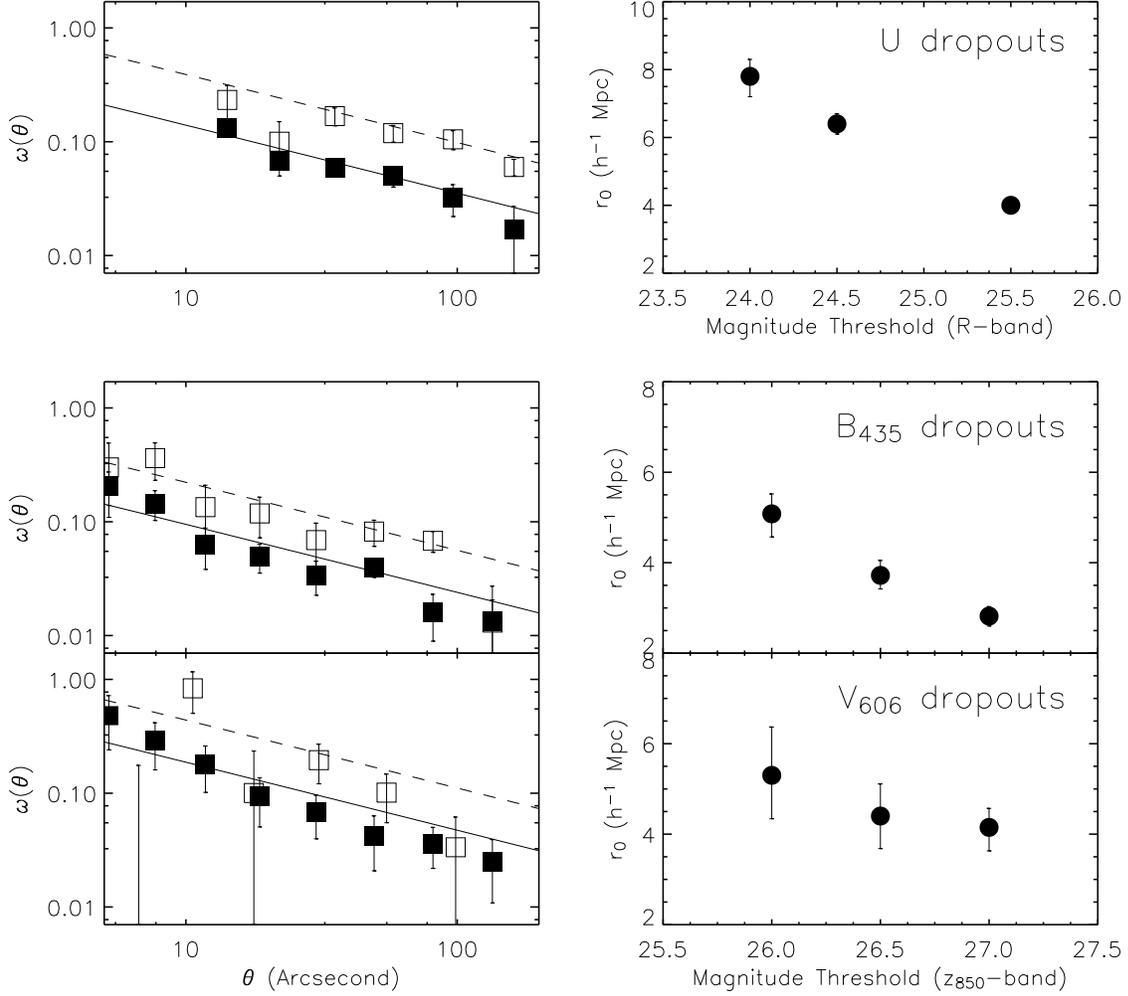}
\caption{Left panels: The angular correlation function \wth\ of the \wu,
\wb, \wv--band dropouts (from top). The filled symbols show the measures from
the full sample (ACS--based: \wz $ \leq 27.0$, ground--based: $R \leq
25.5$) while the open symbols show results from the brightest sub--sample (ACS: \wz $ \leq
26.0$, ground--based: $R \leq 24.0$) together with the best--fit power--law
when the slope is forced to be $\beta = 0.6$. The correlation amplitude,
$A_w$, increases by a factor of more than 2 for the brighter groups in each
case. Right panels: The real--space correlation length $r_0$ is shown as a
function of the magnitude thresholds used to define each sample. Note that
the top--right panel is $R$--based ($R$ $\leq$ 24.0, 24.5 and 25.5), whereas the others
are \wz--based (\wz $\leq $ 26.0, 26.5 and 27.0). }
\end{figure}

\begin{figure}
\epsscale{1.0}
\plotone{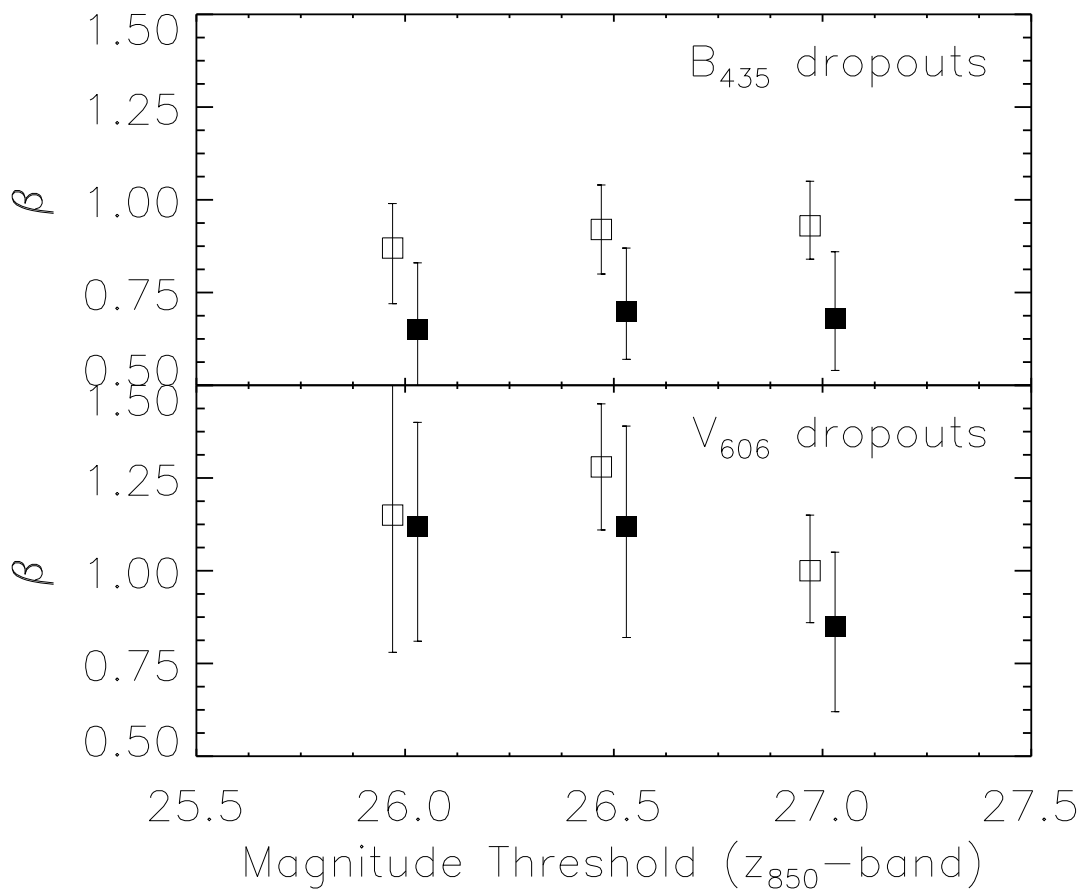}
\caption{The steepening of the correlation slope on small scales: both panels show
the best--fit slope of the correlation function for samples with a given
apparent magnitude threshold. The filled (open) symbols show the slope $\beta$
when fitted to a single power--law without (with) the observational measures
at angular separations of $\theta < 10$ arcsec. This is due to an
excess number of physically close galaxy--pairs 
causing a significant departure from a pure power--law at
small separations.}
\end{figure}

\begin{figure}
\epsscale{1.0}
\plotone{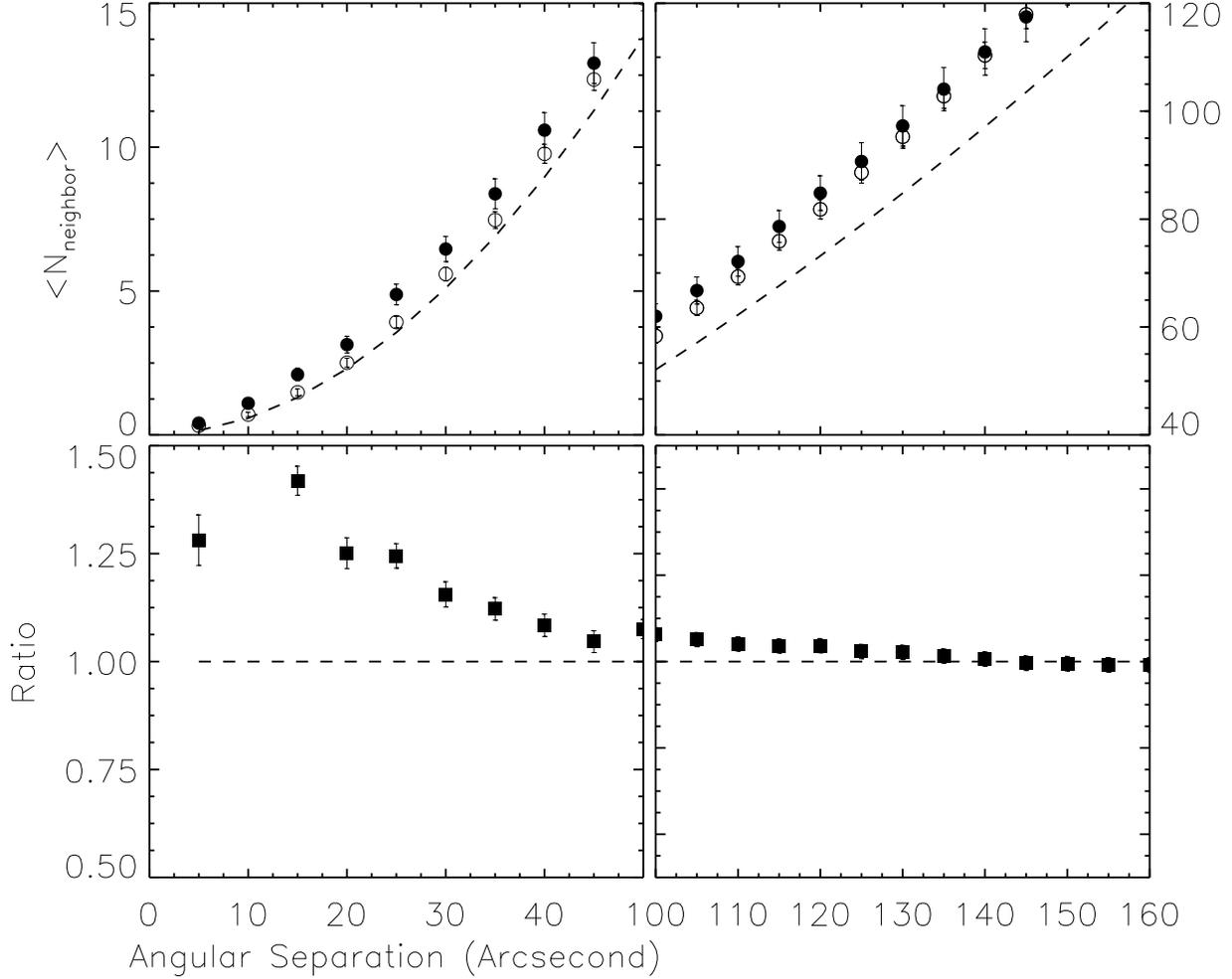}
\caption{The top panels show the average number of faint LBG neighbors (\wz $
\geq m^*+0.5$; $m^* = 24.7$ at $z\sim4$) as a function of angular separation around the two bright groups,
\wz $<24.3$ (filled) and $m^* \leq$ \wz $ \leq 25.0$ (open), together with
what is expected for purely random distributions ($w = 0$), shown as the dashed
lines. The error bars represent Poisson errors. The bottom panels show the
ratio of the two quantities shown in the top panels. Taking the ratio of the two
essentially removes the effect from the projected galaxy pairs, providing a better
indication of the true excess due to physical pairs. The galaxies that belong
to the brightest group (\wz $<24.3$), on average, have more faint
LBG neighbors (3$\sigma$ significance) than the other group ($m^* \leq
$ \wz $\leq 25.0$). 
This effect is most pronounced at $0<\theta<30$ arcsec. 
The ratio slowly drops down to unity on larger scales. 
This is consistent with
the halo sub--structure interpretation because when the angular separation is
comparable to the typical distance between nearby halos, $\langle
N_{neighbor}\rangle$ is dominated by large--scale clustering rather than halo
sub--structure.}
\end{figure}

\begin{figure}
\epsscale{1.0}
\plottwo{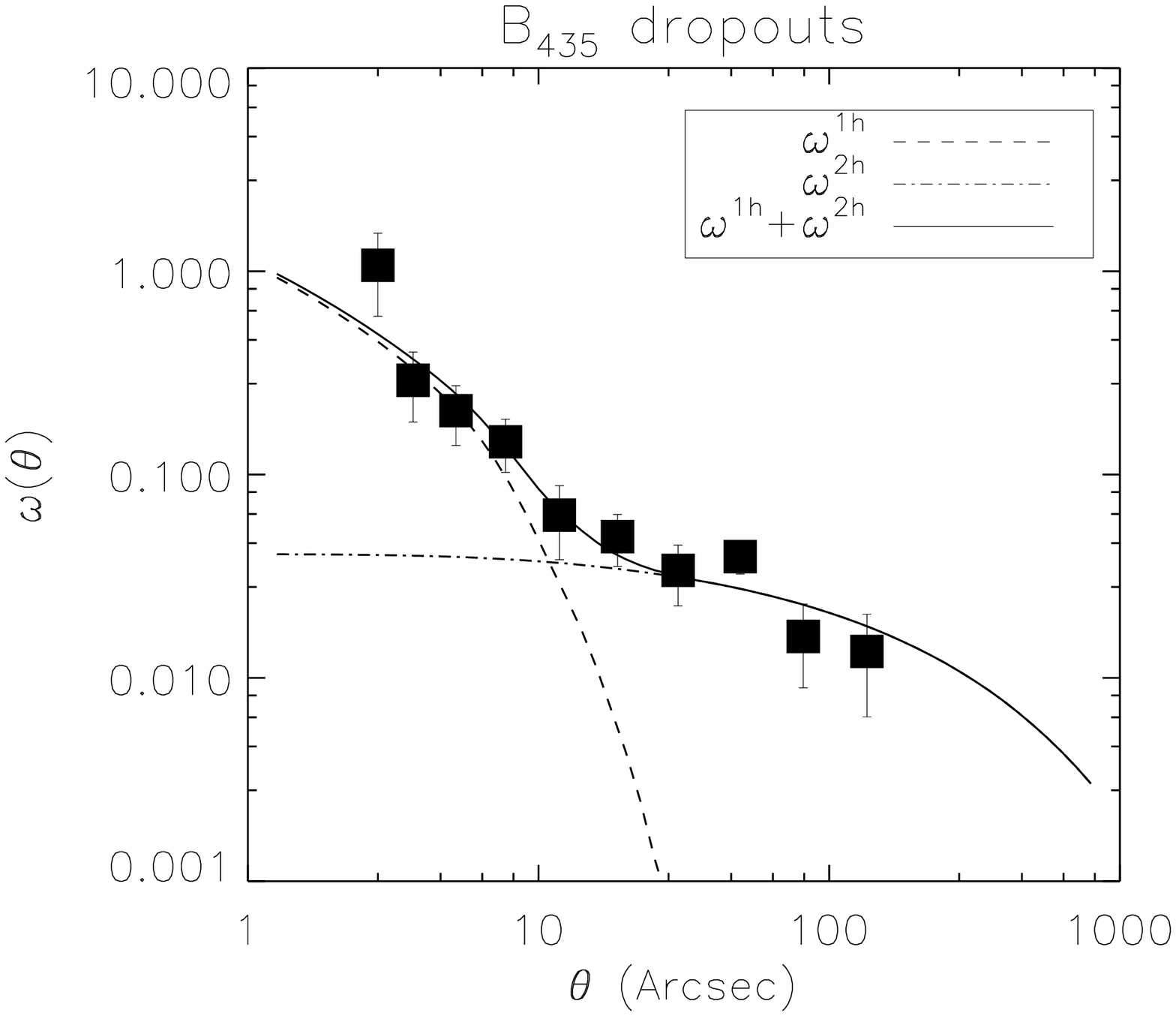}{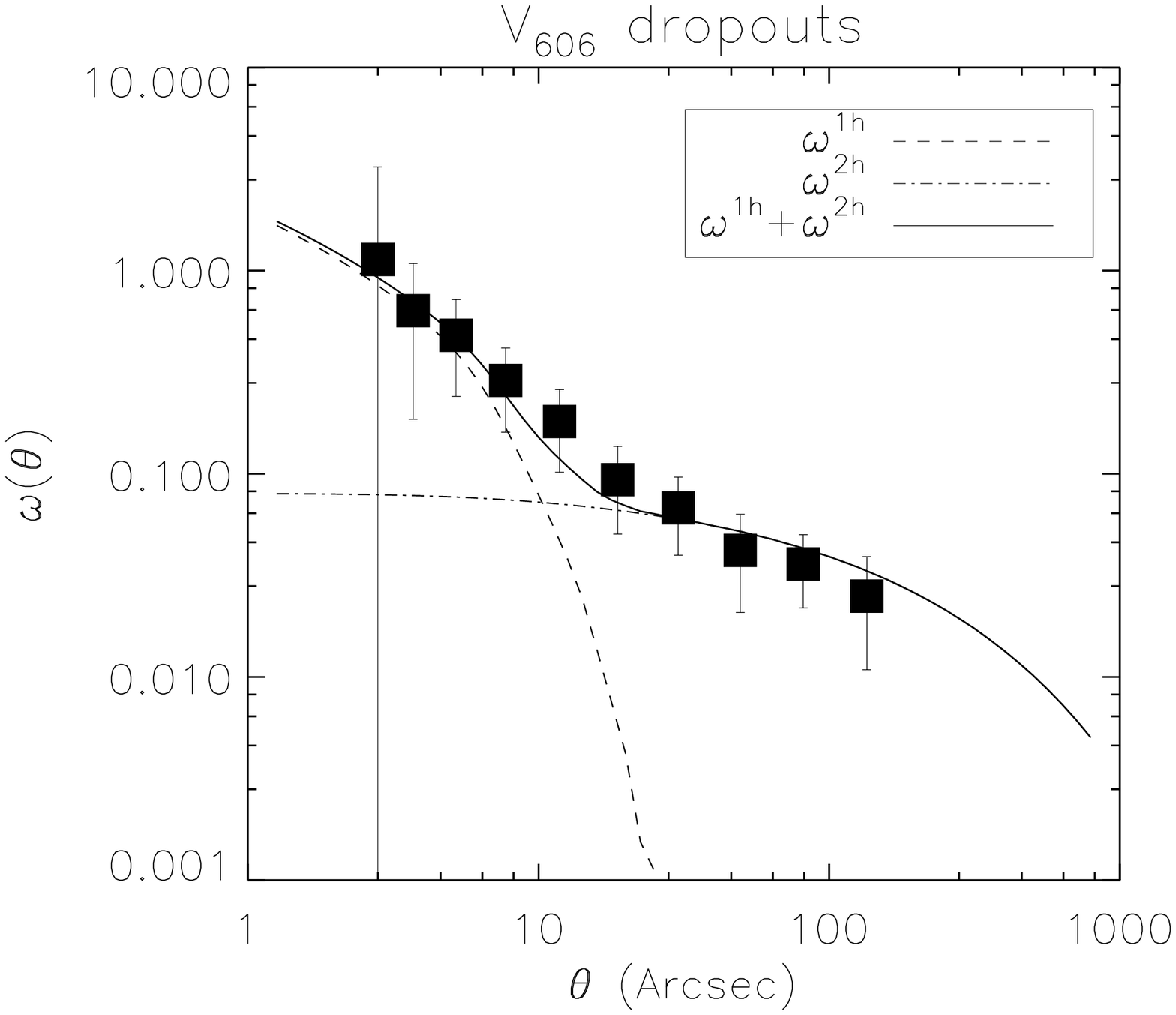}
\caption{The two--component fit with the best--fit HOD parameters is shown together with
the observed angular correlation function. The dashed line shows the one--halo contribution and
the dashed--dot line shows the two--halo contribution to the final
\wth. The solid line is the sum of the two contributions.
The best--fit HOD parameters are $M_{min} = 7 \times
10^{10} h^{-1}M_\sun$, $M_1 = 1.3 \times 10^{12} h^{-1}M_\sun$ and $\alpha$ =
0.65. Right: the \wv--band dropouts: The best--fit HOD parameters are $M_{min}
= 5 \times 10^{10} h^{-1}M_{\sun}$, $M_1 = 1.0 \times 10^{12} h^{-1}M_{\sun}$ and $\alpha$
= 0.80.}
\end{figure}
\clearpage

\begin{figure}
\epsscale{1.0}
\plottwo{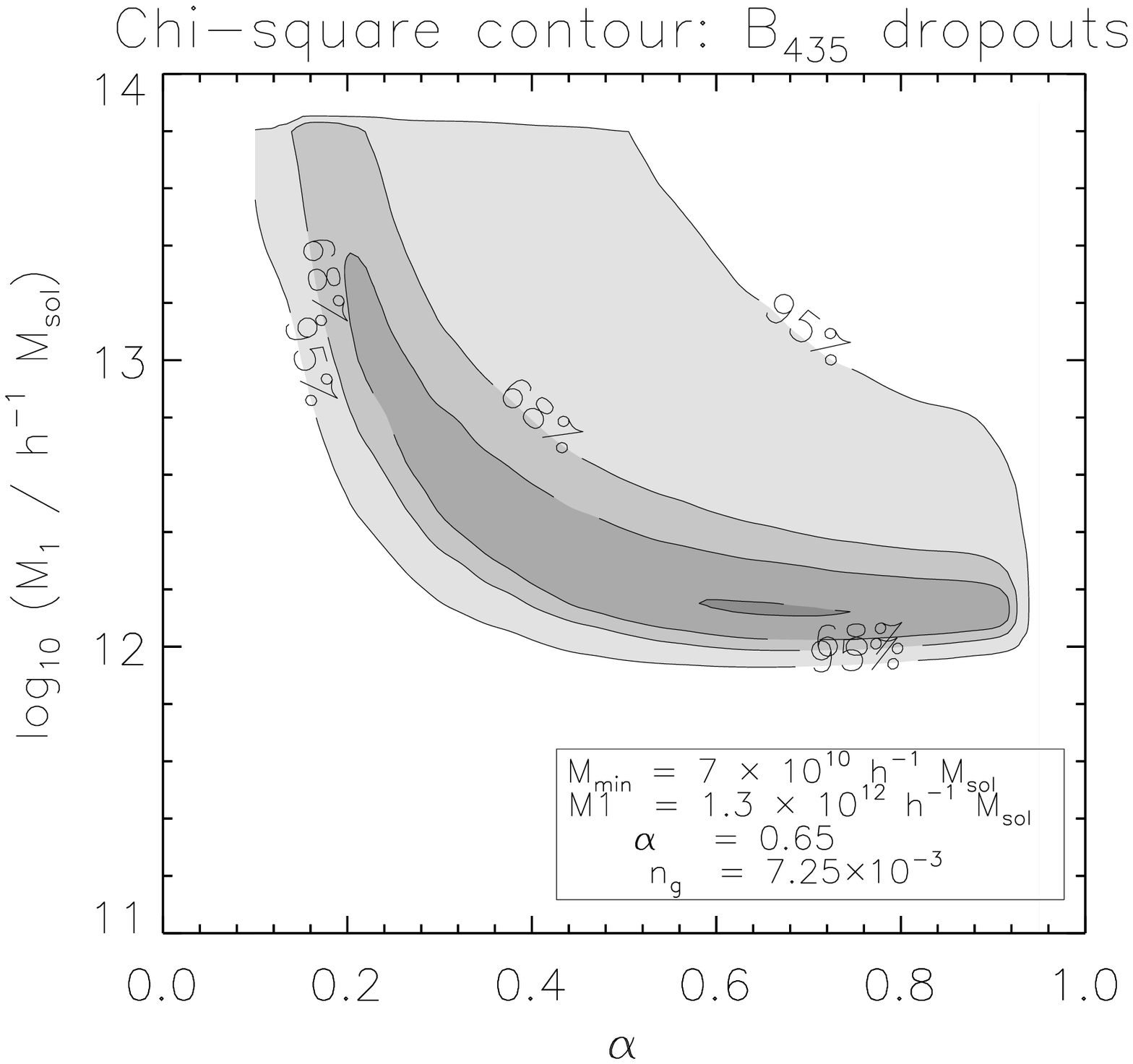}{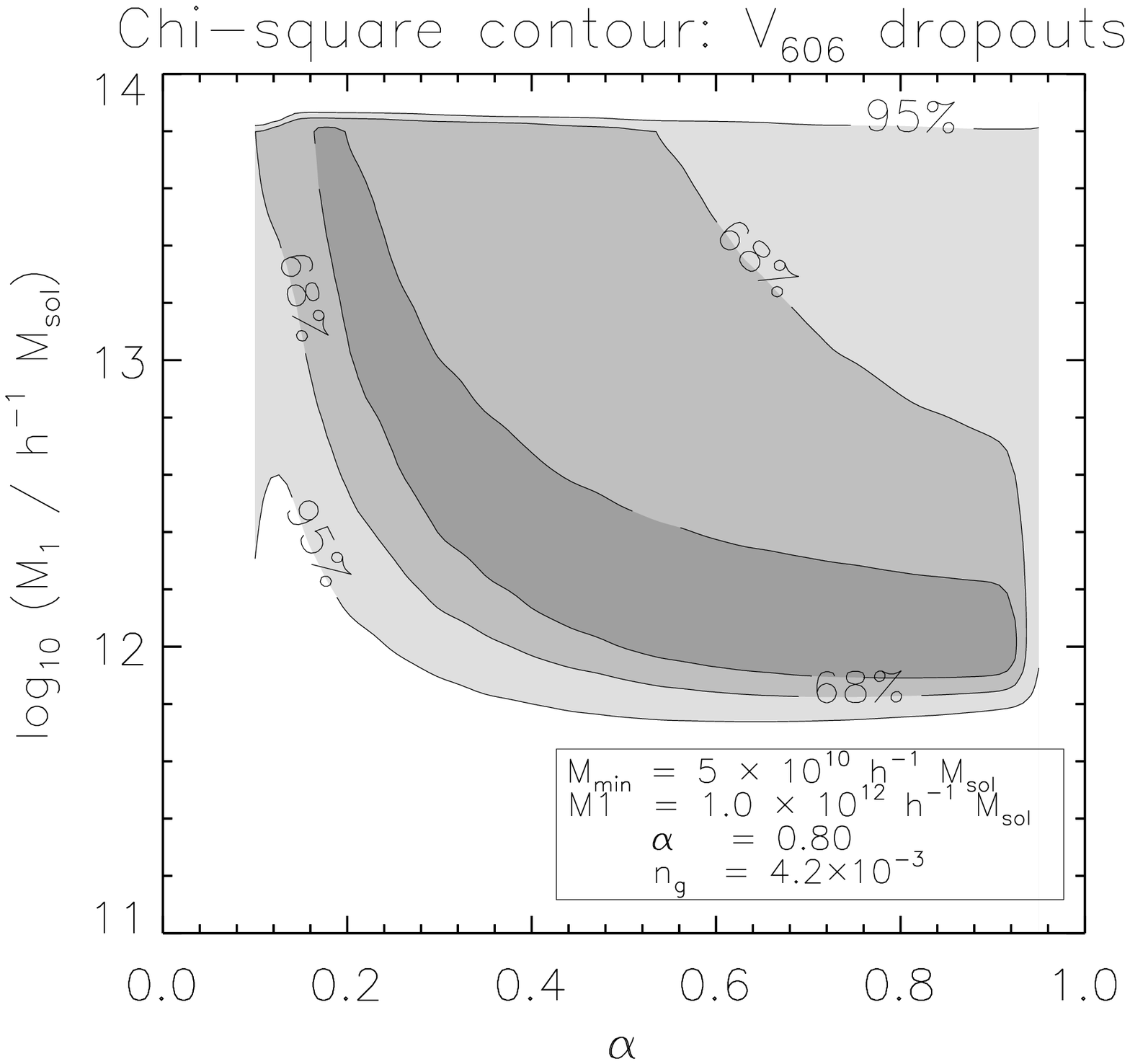}
\caption{The $\chi^2$ contour map of the HOD fits overlaid with confidence intervals for the
\wb\ (left) and \wv--band (right) dropouts: $M_1$ vs. $\alpha$. The contour
lines show $\Delta\chi^2$ = 2.3 and 6.2, corresponding to 68\% and 95\%
confidence levels. Boxes on the bottom right corners show the best--fit HOD
parameter values and the galaxy number density.}
\end{figure}

\clearpage

\begin{figure}
\epsscale{1.0}
\plottwo{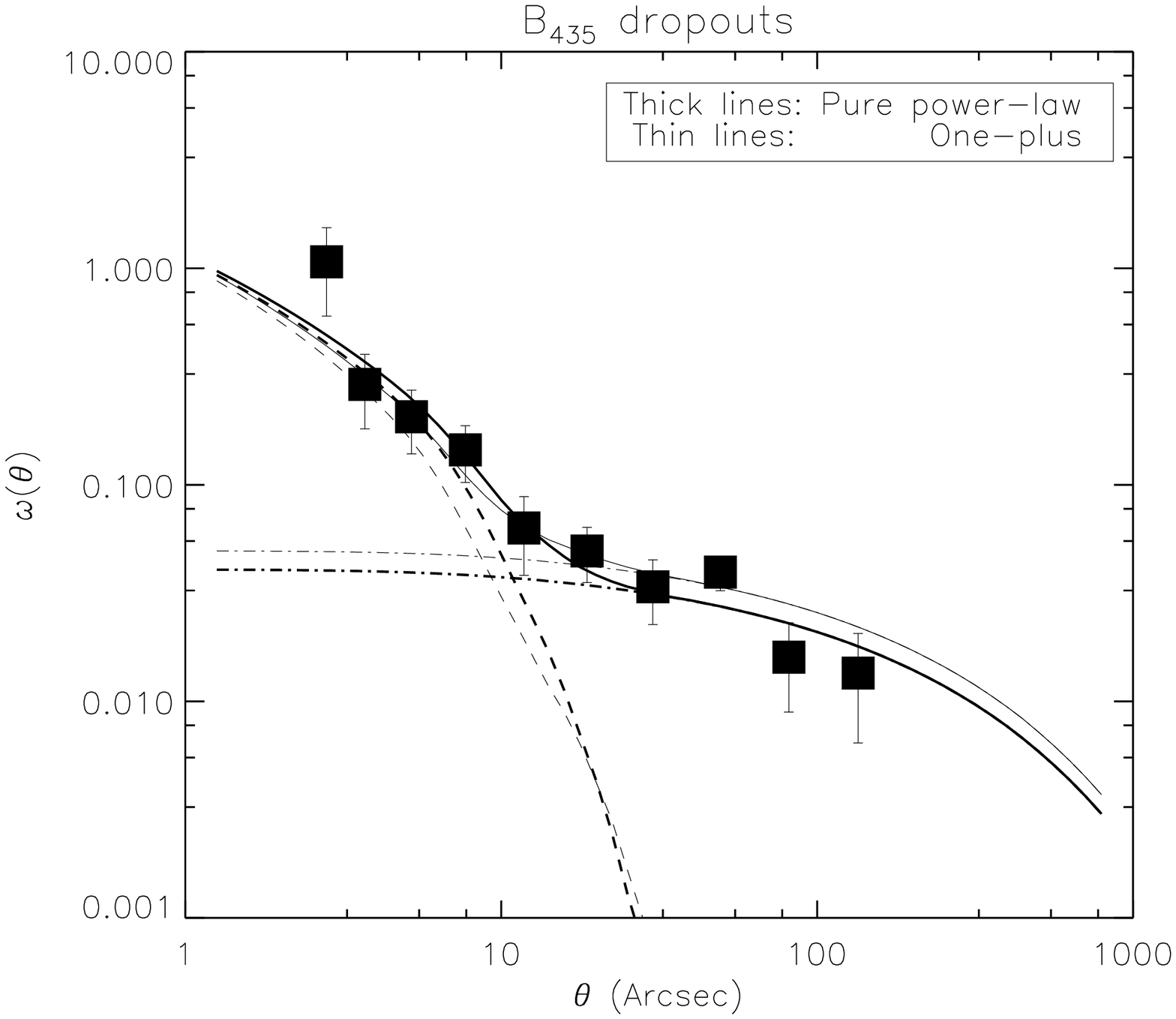}{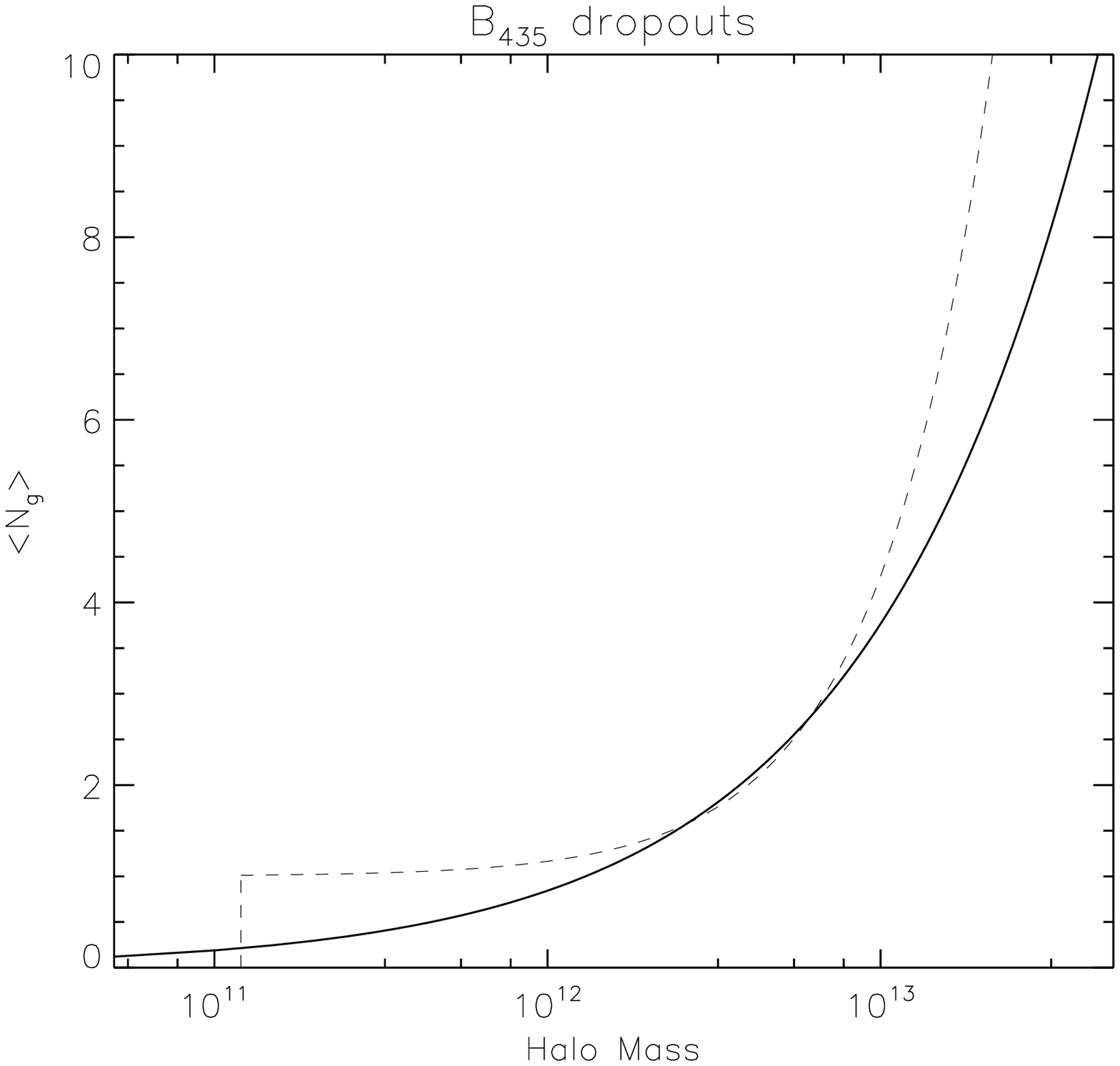}

\caption{Left panel: The two--component fit of the ``pure power--law'' (thick
lines) and ``one--plus'' (thin lines) model is shown together with the
observed angular correlation function of the \wb--band dropouts. The two
models describe the observed correlation function equally well ($\chi^2_{1+}=0.97$
and $\chi^2_{ppl}=0.91$). Right panel: the mean occupation number $\langle N_g
\rangle$ predicted from the two best--fit HOD models is shown as a function of
mass.  Halo mass is in units of $h^{-1} M_\sun$.  The two models predict
similar $\langle N_g \rangle$ values in the mass range of $10^{12}$ --
$10^{13}$ $h^{-1}M_\sun$, to which most of the massive halos ``observed'' in
the data most likely belong. This implies that our estimate is very robust
regardless of the employed HOD model in the high--end of the halo mass
function.  However, at the mass scale of $M>10^{13} h^{-1} M_\sun$, $\langle
N_g \rangle $ from the ``one--plus'' model largely exceeds the other. On the
other hand, at $M<10^{11} h^{-1}M_\sun$, the ``pure power--law'' model implies
a larger number of low--mass halos in the samples, while the ``one--plus''
model is truncated (by definition; see text for further discussion).  
}

\end{figure}

\begin{figure}
\epsscale{1.0}
\plotone{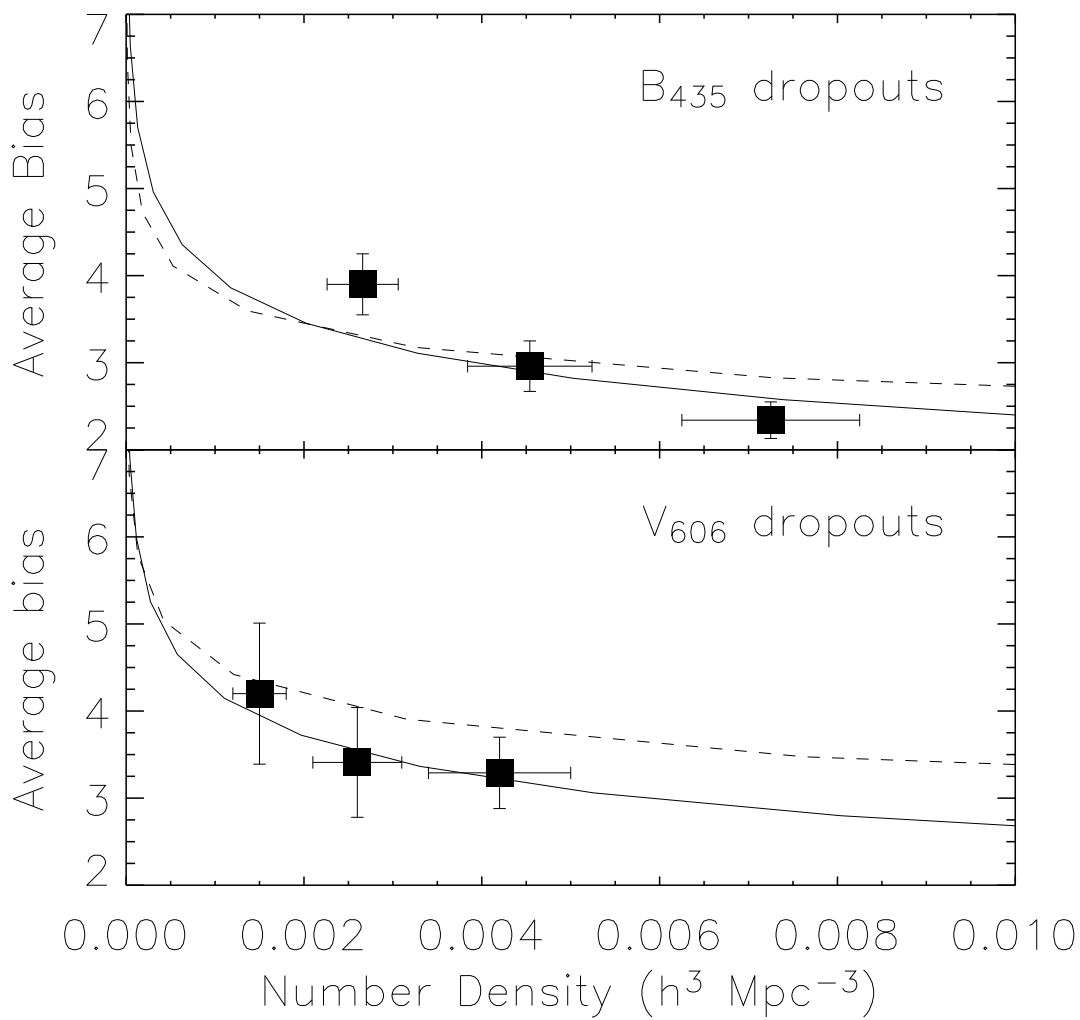}
\caption{The average bias vs the galaxy (or halo) abundance: The data points and the solid lines represent the
average galaxy bias from the correlation measures and the predictions as a
function of galaxy number density, respectively. The average bias is computed as linear bias
(Sheth \& Tormen 1999) weighted by the best--fit HOD to account for the galaxy
multiplicity. The dashed lines show the average halo bias as a function of halo
number density.  If there were one--to--one correspondence between halos and
galaxies, then the two would coincide. }
\end{figure}

\begin{figure}
\epsscale{1.0}
\plotone{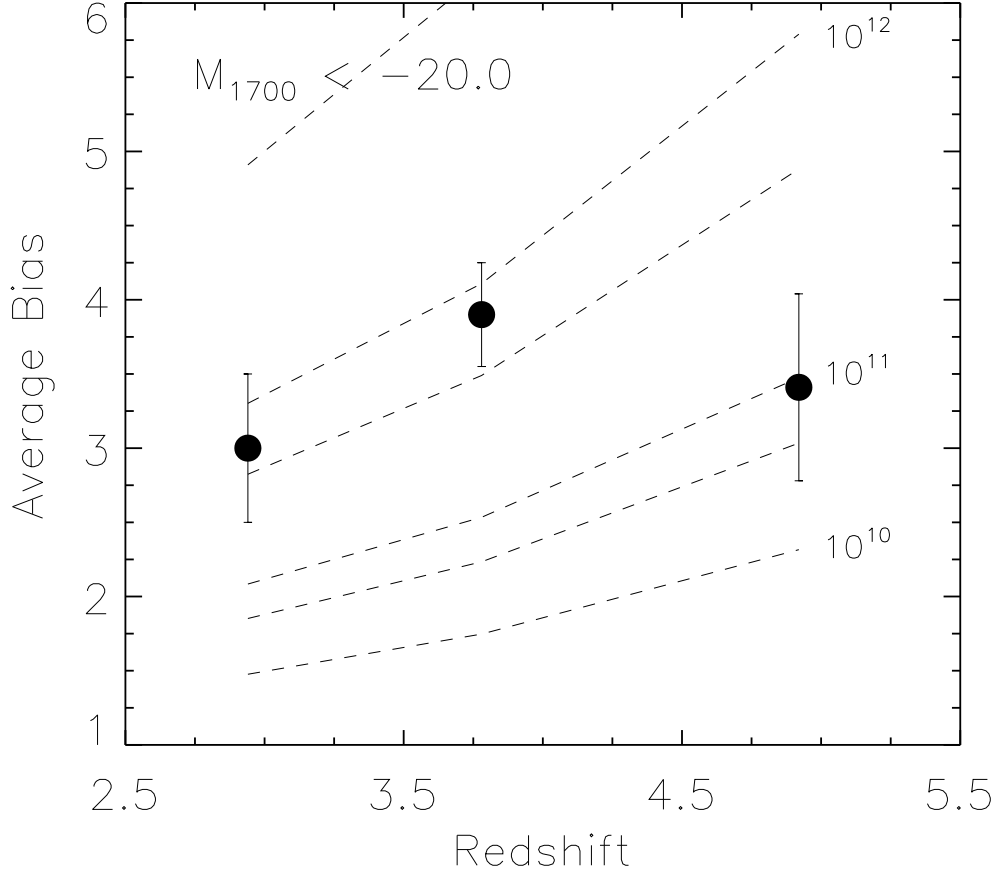}
\caption{The evolution of the average bias as a function of redshift: the data points
indicate the average bias implied for the \wu, \wb\ and \wv--band dropouts 
at the fixed absolute luminosity threshold $M_{1700} \leq -20.0$, which corresponds to
${\cal R} \leq 25.5$ at {\it z} $\sim$ 3. The dashed lines show the predicted
linear bias from the Sheth \& Tormen (1999) model for halo mass $M_{halo}$ $\geq$
$M_{min}$ = $10^{10},5\times10^{10},10^{11},5\times10^{11},10^{12}$ and
$5\times10^{12}$ $h^{-1} M_{\sun}$ (from bottom). At $z \sim$ 3 and 4, the observed
galaxies are hosted by dark halos of similar mass
($5\times10^{11}h^{-1}M_{\sun}$ -- $10^{12}$ $h^{-1}M_{\sun}$). At $z \sim$
5, the observed clustering strength of the galaxies implies that they are
hosted by less massive halos, $M_{halo} \sim 10^{11} h^{-1}M_{\sun}$. This suggests
that star--formation may have been more efficient at {\it z} $\sim$ 5. }
\end{figure}

\clearpage

\begin{table}
\begin{center}
\caption{The space--based data set\label{tbl-1}}
\begin{tabular}{crrrrrrrrrr}
\\
\tableline\tableline
Field & A\tablenotemark{a} & $\sigma(B_{435})$\tablenotemark{b} & $\sigma(V_{606})$\tablenotemark{b} & $\sigma(i_{775})$\tablenotemark{b} & $\sigma(z_{850})$\tablenotemark{b} & $N_{LBG,4}$\tablenotemark{c} & $N_{LBG,5}$\tablenotemark{c}\\
\tableline
GOODS--N&160.0& 29.00 & 29.08 & 28.33 & 28.09 & 1169 & 502 \\
GOODS--S&160.0& 28.88 & 29.05 & 28.34 & 28.06 & 1294 & 376\\
\tableline
\end{tabular}
\tablenotetext{a}{Survey area, in units of arcmin$^2$}
\tablenotetext{b}{1 $\sigma$ surface brightness fluctuations within 1 arcsec$^2$ apertures}
\tablenotetext{c}{Number of \wb/\wv--band dropout Lyman--break galaxies}
\end{center}
\end{table}

\begin{table}
\begin{center}
\caption{The ground--based data set\label{tbl-2}}
\begin{tabular}{crrrrrrrrrr}
\\
\tableline\tableline
Field & A & $\sigma(U)$\tablenotemark{a} & $\sigma$($B$)\tablenotemark{a} & $\sigma$($R$)\tablenotemark{a} & $N_U$\tablenotemark{b} \\
\tableline
extended GOODS--S&1200& 29.42 & 29.75 & 28.95 & 1609 \\
\tableline
\end{tabular}
\tablenotetext{a}{1 $\sigma$ surface brightness fluctuations within 1 arcsec$^2$ apertures}
\tablenotetext{b}{Number of \wu--band dropout Lyman--break galaxies ($R \leq 25.5$)}
\end{center}
\end{table}

\begin{table}
\begin{center}
\caption{The LBG abundance, the angular correlation function and the inferred correlation
lengths \label{tbl-3}}
\begin{tabular}{crrrrrrrrrrr}
\\
\tableline\tableline
Flavor & Sample & $n_g(\times 10^{-3})$ & $A_{w,0}$ & 
\multicolumn{1}{c}{$\beta_0$} & {$r_{0,0}$} & 
\multicolumn{1}{c}{$A_{w,1}$} & $\beta_1$ & $r_{0,1}$\\
\tableline
\wu&$R\leq$ 25.5&$3.3\pm1.0$&$0.56^{+0.04}_{-0.04}$&0.6&$4.1^{+0.1}_{-0.2}$&$0.52^{+0.28}_{-0.20}$&$0.63^{+0.11}_{-0.12}$&$4.0^{+0.2}_{-0.2}$\\
\wu&$R\leq$ 24.5&$0.7\pm0.3$&$1.16^{+0.06}_{-0.08}$&0.6&$6.5^{+0.2}_{-0.3}$&$0.54^{+0.23}_{-0.16}$&$0.44^{+0.08}_{-0.08}$&$6.4^{+0.3}_{-0.3}$\\
\wu&$R\leq$ 24.0&$0.2\pm0.1$&$1.58^{+0.14}_{-0.17}$&0.6&$7.8^{+0.4}_{-0.5}$&$0.84^{+0.68}_{-0.32}$&$0.52^{+0.12}_{-0.11}$&$7.8^{+0.5}_{-0.6}$\\
\wb&\wz $ \leq$ 27.0 & $7.3\pm1.0$ &$0.38^{+0.04}_{-0.05}$ & 0.6 & $2.9^{+0.2}_{-0.2} $ & $0.42^{+0.32}_{-0.26}$ & $0.69^{+0.16}_{-0.15}$ & $2.8^{+0.2}_{-0.2}$ \\
\wb&\wz $ \leq$ 26.5 & $4.5\pm0.7$ & $0.60^{+0.10}_{-0.06}$ &0.6 & $3.9^{+0.3}_{-0.3} $ & $0.75^{+0.56}_{-0.37}$ & $0.70^{+0.17}_{-0.13}$ & $3.7^{+0.3}_{-0.3}$\\
\wb&\wz $ \leq$ 26.0 & $2.7\pm0.4$ & $0.99^{+0.14}_{-0.13}$  & 0.6 & $5.3^{+0.4}_{-0.5}$ & $0.64^{+0.98}_{-0.45}$ & $0.64^{+0.19}_{-0.19}$ & $5.1^{+0.4}_{-0.5}$\\
\wv&\wz $ \leq$ 27.0 & $4.2\pm0.8$ &$0.76^{+0.13}_{-0.15}$ & 0.6& $4.4^{+0.5}_{-0.5} $ & $1.00^{+1.68}_{-0.75}$ & $0.85^{+0.20}_{-0.23}$ & $4.2^{+0.4}_{-0.5}$\\
\wv&\wz $ \leq$ 26.5 & $2.6\pm0.5$ & $1.12^{+0.34}_{-0.25}$ & 0.6 & $5.8^{+1.1}_{-0.8} $ & $2.08^{+8.80}_{-1.76}$ & $1.10^{+0.30}_{-0.27}$ & $4.4^{+0.7}_{-0.7}$ \\
\wv&\wz $ \leq$ 26.0 & $1.5\pm0.3$ & $1.70^{+0.42}_{-0.37}$ & 0.6 & $7.5^{+1.1}_{-1.0}$ & $3.88^{+12.89}_{-3.46}$ & $1.10^{+0.31}_{-0.29}$ & $5.3^{+1.1}_{-1.0}$\\
\tableline
\end{tabular}
\tablecomments{The number density ($n_g$) and the comoving correlation lengths ($r_0$)
are in units of $h^3$ Mpc$^{-3}$ and $h^{-1}$ Mpc, respectively. The quantities
with the secondary subscript ``0'' are when the slope is fixed to a fiducial value
$\beta$ = 0.6, those with the subscript ``1'' are when both $\beta$ and $A_w$ are
allowed to vary. }
\end{center}
\end{table}

\begin{table}
\begin{center}
\caption{The best--fit HOD parameters\tablenotemark{a}, 
the ratio of galaxy number density to halo number density $\langle N_g \rangle_M$,
and the average halo mass $\langle M \rangle$\tablenotemark{b} for different luminosity thresholds \label{tbl-4}}
\begin{tabular}{crrrrrrrrrrr}
\\
\tableline\tableline
Flavor & Sample & $M_1\tablenotemark{b,c}$ & $\alpha$\tablenotemark{c} &
\multicolumn{1}{c}{$\langle N_g \rangle_M$} & $\langle M \rangle$\tablenotemark{b} \\
\tableline
\wb&\wz $ \leq$ 27.0&1.3$\times 10^{12}$&0.65&0.31$\pm$0.14&(4$\pm$1)$\times10^{11}$\\
\wb&\wz $ \leq$ 26.5&1.3$\times 10^{12}$&0.65&0.38$\pm$0.13&(6$\pm$1)$\times10^{11}$\\
\wb&\wz $ \leq$ 26.0&1.3$\times 10^{12}$&0.65&0.49$\pm$0.13&(8$\pm$2)$\times10^{11}$\\
\wv&\wz $ \leq$ 27.0&1.0$\times 10^{12}$&0.80&0.20$\pm$0.18&(2$\pm$1)$\times10^{11}$\\
\wv&\wz $ \leq$ 26.5&1.0$\times 10^{12}$&0.80&0.24$\pm$0.19&(3$\pm$1)$\times10^{11}$\\
\wv&\wz $ \leq$ 26.0&1.0$\times 10^{12}$&0.80&0.30$\pm$0.21&(4$\pm$2)$\times10^{11}$\\
\tableline
\end{tabular}
\tablenotetext{a}{Using the ``pure power--law'' model}
\tablenotetext{b}{All masses are in units of $h^{-1}M_\sun$ }
\tablenotetext{c}{Note that the same HOD parameters, $M_1$ and $\alpha$, constrained from the full sample, were used for all sub--samples}
\end{center}
\end{table}


\begin{thebibliography}{}

\bibitem[Adelberger et al.(1998)]{adel98} Adelberger, K. et al. 1998, \apj,
505, 18
\bibitem[Adelberger \& Steidel(2000)]{as00} Adelberger, K. \& Steidel,
C. 2000, \apj, 544, 218
\bibitem[Adelberger et al.(2005)]{adel05} Adelberger, K. et al. 2005, \apj, 619, 697
\bibitem[Allen et al.(2005)]{allen05} Allen, P. D. et al. 2005, \mnras, 360, 1244
\bibitem[Bagla (1998)]{bagla98} Bagla, J. S. 1998, \mnras, 297, 251
\bibitem[Benoist et al. (1996)]{benoist96} Benoist, C. et al. 1996, \apj, 472, 452
\bibitem[Berlind \& Weinberg(2002)]{bw02} Berlind, A. A. \& Weinberg, D. H. 2002, \apj, 575, 587
\bibitem[Berlind et al.(2003)]{ber03} Berlind, A. A. et al. 2003, \apj, 593, 1
\bibitem[Bertin \& Arnouts (1996)]{ba96} Bertin, E. \& Arnouts, S. 1996, \aaps, 117, 393
\bibitem[Bullock et al.(2001)]{bull01} Bullock, J. et al. 2001, \mnras, 321,
559
\bibitem[Bullock et al.(2002)]{bull02} Bullock, J. et al. 2002, \mnras, 329,
246
\bibitem[Chapman et al.(2003)]{chapman03} Chapman, S. et al. 2003, \nat, 422,
695
\bibitem[Cole \& Kaiser(1989)]{cole89} Cole, S. \& Kaiser, N. 1989, \mnras,
237, 1127
\bibitem[Daddi et al.(2003)]{daddi03} Daddi, E. et al. 2003, \apj, 588, 50
\bibitem[Dickinson et al.(2004)]{med04} Dickinson, M. E. et al. 2004, \apj, 600, L99
\bibitem[van Dokkum et al. (2003)]{vd03} van Dokkum, P. et al. 2003, \apjl, 587, 83
\bibitem[Efstathiou (2000)]{efstathiou00} Efstathiou, G. 2000, \mnras, 317, 697
\bibitem[Ferguson et al.(2004)]{ferg04} Ferguson, H. et al. 2004, \apjl, 600,
L107
\bibitem[Foucaud et al. (2003]{foucaud03} Foucaud, S. et al. 2003 \aap, 409, 835
\bibitem[Franx et al.(2003)]{franx03} Franx, M. et al. 2003, \apjl, 587, L70
\bibitem[Giavalisco et al.(1998)]{mauro98} Giavalisco, M. et al. 1998, \apj,
503, 543
\bibitem[Giavalisco(2002)]{mauro02} Giavalisco, M. 2002, \araa , 40, 579
\bibitem[Giavalisco \& Dickinson(2001)]{GD01} Giavalisco, M. \& Dickinson, M. E.
2001, \apj, 550, 177
\bibitem[Giavalisco et al.(2004a)]{mauro04a} Giavalisco, M. et al. 2004a, \apjl, 600, L93
\bibitem[Giavalisco et al.(2004b)]{mauro04b} Giavalisco, M. et al. 2004b, \apjl, 600, L103
\bibitem[Giavalisco (2005)]{mauro05a} Giavalisco, M. 2005, {\it proceedings from Wide--Field Imaging from Space} eds. Tim McKay, Andy Fruchter, and Eric Linder
\bibitem[Giavalisco et al.(2005)]{mauro05b} M. Giavalisco et al., in preparation 
\bibitem[Gunn \& Oke (1975)]{go75} Gunn, J. E. \& Oke, J. B. 1975, \apj, 195, 225
\bibitem[Guzik \& Seljak (2002)]{gs02} Guzik, J. \& Seljak, U. 2002, \mnras, 335, 311
\bibitem[Hamana et al.(2004)]{hamana04} Hamana, T. et al. 2004, \mnras, 347, 813
\bibitem[Jing et al.(1998)]{jing98} Jing, Y. P., Mo, H. J. and Boerner, G. 1998, \apj, 494, 1
\bibitem[Kravtsov et al.(2004)]{krav04} Kravtsov, A. et al. 2004, \apj, 609, 35
\bibitem[Landy \& Szalay(1993)]{sz93} Landy, S. \& Szalay, A. 1993, \apj, 412, 64
\bibitem[Lee et al. (2005)]{lee05} K.-S. Lee et al., in preparation
\bibitem[Ling, Barrow \& Frenk (1986)]{ling86} Ling, E. N., Barrow, J. D. \& Frenk, C. S. 1986, \mnras, 223, 21
\bibitem[Madau et al.(1995)]{madau95} Madau, P. et al. 1995, \apj, 441, 18
\bibitem[Madau et al.(1996)]{madau96} Madau, P. et al. 1996, \mnras, 283, 1388
\bibitem[Magliocchetti et al. (2003)]{maglio03} Magliocchetti, M. et al. 2003, \mnras, 346, 186
\bibitem[Miley et al.(2004)]{miley04} Miley, G. et al. 2004, \nat, 427, 47
\bibitem[Mo \& White(1996)]{MW96} Mo, H. J., \& White, S. D. M. 1996, \mnras,
282, 347
\bibitem[Navarro, Frenk \& White(1997)]{nfw97} Navarro, J. F., Frenk, C. S. \&
White, S. D. M. 1997, \apj, 462, 563
\bibitem[Norberg et al.(2001)]{nord01} Norberg, P. et al. 2001, \mnras, 328, 64
\bibitem[Ouchi et al.(2004)]{ouchi04} Ouchi, M. et al. 2004, \apj, 611, 685
\bibitem[Ouchi et al.(2005a)]{ouchi05a} Ouchi, M. et al. 2005a, \apj, 620, L1
\bibitem[Ouchi et al.(2005b)]{ouchi05b} Ouchi, M. et al. 2005b, submitted to \apj
\bibitem[Peacock \& Dodds (1996)]{pd96} Peacock, J. A. \& Dodds, S. J. 1996, \mnras, 280, 19
\bibitem[Peebles(1980)]{peeb80} Peebles, P. J. E. 1980, The Large--Scale
Structure of the Universe (Princeton Univ. Press)
\bibitem[Porciani \& Giavalisco (2002)]{cris02} Porciani, C. \& Giavalisco, M. 2002, \apj, 565, 24
\bibitem[Porciani et al.(2004)]{cris04} Porciani, C. et al. 2004, \mnras, 355, 1010
Recipes (Cambridge: Cambridge Univ. Press)
\bibitem[Riess et al.(2004)]{riess04} Riess, A. et al. 2004, \apjl, 600, L163
\bibitem[Roche \& Eales (1999)]{roche99} Roche, N. \& Eales, S. 1999, \mnras,
307, 703
\bibitem[Sheth \& Tormen(1999)]{st99} Sheth, R. K. \& Tormen, G. 1999, \mnras, 308, 119
\bibitem[Sheth \& Tormen(2002)]{st02} Sheth, R. K. \& Tormen, G. 2002, \mnras, 329, 61 
\bibitem[Scherrer \& Bertschinger(1991)]{scherrer91} Scherrer, R. J. \& Bertschinger, E. 1991, \apj, 381, 349
\bibitem[Seljak(2000)]{seljak00} Seljak, U. 2000, \mnras, 318, 203
\bibitem[Shapley et al.(2004)]{shapley04} Shapley, A. E. et al. 2004, \apj, 612, 108
\bibitem[Silk (1997)]{silk97} Silk, J. 1997, \mnras, 481, 703
\bibitem[Somerville et al.(2004)]{somer04} Somerville, R. et al. 2004, \apjl,
600, L171
\bibitem[Steidel et al.(1995)]{steidel95} Steidel, C. C. et al. 1995, \aj,
110, 2519
\bibitem[Steidel et al.(1998)]{steidel98} Steidel, C. et al. 1998, \apj, 492,
428
\bibitem[Steidel et al.(1999)]{steidel99} Steidel, C. et al. 1999, \apj, 519,
1
\bibitem[Steidel \& Hamilton(1993)]{sh93} Steidel, C. C. \& Hamilton, D. 1993,
\apj, 105, 2017
\bibitem[Steidel et al.(2003)]{steidel03} Steidel, C. C. et al. 2003, \apj, 592, 728
\bibitem[van den Bosch et al.(2003)]{vdb03} van den Bosch, F. C., Yang, X., Mo, H. J. 2003, \mnras, 340, 771
\bibitem[Vanzella et al. (2005)]{vanzella05a} Vanzella, E. et al. 2005, \aap, 434, 53
\bibitem[Vanzella et al. (2005)]{vanzella05b} E. Vanzella et al., in preparation
\bibitem[Wechsler et al. (2001)]{wechsler01} Wechsler, R. et al. 2001, \apj,
554, 85
\bibitem[White \& Reese(1978)]{wr78} White, S. D. M. \& Reese, M. 1978,
\mnras, 183, 341
\bibitem[Yang et al.(2003)]{yang03} Yang, X., Mo, H. J., van den Bosch, F. C. 2003,  \mnras, 339, 1057
\bibitem[Zehavi et al.(2002)]{zehavi02} Zehavi, I. et al. 2002, \apj, 571, 172
\bibitem[Zehavi et al.(2004a)]{zehavi04a} Zehavi, I. et al. 2004a, \apj, 608, 16
\bibitem[Zehavi et al.(2004b)]{zehavi04b} Zehavi, I. et al. 2004b, astro-ph/0408569
\bibitem[Zheng et al.(2005)]{zheng05} Zheng, Z. et al. 2005, \apj, 633, 791
\end{thebibliography}
\end{document}